\documentclass[a4paper]{article} 
\usepackage{lipsum}
\usepackage{url}
\usepackage{mathpazo}
\usepackage[T1]{fontenc}
\usepackage[utf8]{inputenc}
\usepackage[english]{babel}

\usepackage[left=3cm,top=3cm,right=3cm,bottom=3cm]{geometry}

\usepackage{natbib}
\usepackage{amsmath}
\usepackage{amssymb}
\usepackage{amsthm}
\usepackage{hyperref}
\usepackage{graphicx}
\usepackage{array}
\usepackage{longtable}
\usepackage{verbatimbox}
\usepackage{textcomp}
\usepackage{gensymb}
\usepackage{verbatimbox}
\newcommand\Includegraphics[2][]{\addvbuffer[3pt 0pt]{\includegraphics[#1]{#2}}}
\usepackage[title]{appendix}
\graphicspath{{fig/}}
\usepackage{bm}
\usepackage{upgreek}
\usepackage{mathrsfs}
\usepackage{xcolor}
\usepackage{placeins}

\usepackage{subcaption}
\usepackage[small, sc]{titlesec}

\usepackage{xcolor}
\hypersetup{
colorlinks,
linkcolor={red!50!black},
citecolor={blue!50!black},
urlcolor={blue!80!black}
}
\theoremstyle{plain} 
\newtheorem{thm}{Theorem}

\newtheorem{rem}[thm]{Remark}

\usepackage{mathtools}
\usepackage{pgfplots}
\pgfplotsset{/pgf/number format/use comma,compat=newest}

\renewcommand\epsilon{\varepsilon}
\newcommand{\R}{\mathbb{R}}
\renewcommand{\r}{\mathbf{r}}
\renewcommand{\c}{\mathbf{c}}
\newcommand{\n}{\mathbf{n}}

\newcommand{\abc}{\phi_{abc}}
\newcommand{\pqr}{\phi_{pqr}}
\newcommand{\V}{\mathcal{V}}

\begin{document}
\title{\textsc{Symmetry break in the eight bubble compaction}}

\author{\textsc{Giulia Bevilacqua}$^1$\thanks{\href{mailto:giulia.bevilacqua@polimi.it}{\texttt{giulia.bevilacqua@polimi.it}}}
\bigskip\\
\normalsize$^1$ MOX -- Politecnico di Milano, Piazza Leonardo da Vinci 32, Milano, Italy}
\date{}

\maketitle

\begin{abstract}

Geometry and mechanics have both a relevant role in determining the three-dimensional packing of $8$ bubbles displyaed in a foam structure. 

We assume that the spatial arrangement of bubbles obeys a geometrical principle maximizing the minimum mutual distance between the bubble centroids. The compacted structure is then obtained by radially packing the bubbles under constraint of volume conservation.  We generate a polygonal tiling on the central sphere and peripheral bubbles with both flat and curved interfaces. We verify that the obtained polyhedra is optimal under suitable physical criteria. Finally, we enforce the mechanical balance imposing the constraint of conservation of volume.

We find an anisotropy in the distribution of the field of forces: surface tensions of bubble-bubble interfaces with normal oriented in the circumferential direction of bubbles aggregate are larger than the ones with normal unit vector pointing radially out of the aggregate. We suggest that this mechanical cue is key for the symmetry break of this bubbles configuration.\\


\end{abstract}

\section{Introduction}

Often used for children's enjoyment, soap bubbles are the simplest physical example of a lot of mathematical problems: they are the solution of the minimal surface problem \citep{plateau1869recherches}, they solve a stability problem since their longevity is limited \citep{saye2013multiscale} and  when two or more bubbles cluster together, their configuration obeys a shape optimality problem \citep{bikerman1973formation}. Assembling several bubbles traps pockets of gas in a liquid and results in {\em foam}: the surfactants added to the liquid stabilize the bubbles by reducing the surface tension and by arranging themselves at the liquid/gas interfaces \citep{weaire2001physics}.

Regarding the behaviour of a single soap bubble, everything is known. What is not completely understood is the geometrical and mechanical properties of a cluster of many bubbles, known as a foam. For instance, its optimal rearrangement in space is still matter of debate.

In $2$D there are more results: Hales proved the {\em honeycomb conjecture}, which states that  the partition of the plane into regular hexagons of equal area has least perimeter, {\em i.e.} it minimizes the perimeter fixing the area \citep{hales2001honeycomb}. In this context, some years later, Morgan proved that the optimal ({\em i.e.} minimal) configuration exists for $N$ clusters \citep{morgan2016geometric} and Cox {\em et al.} obtained their numerical visualisations up to $N = 200$ \citep{cox2003minimal}. Due to the non linearity of the problem, in a lot of physical situations, the equilibrium solution is only stable with respect to {\em small displacements}, {\em i.e.} it is not a global minimum of the system. This aspect leads to mechanical instabilities which break the symmetry of the system \citep{brakke2002instabilities,  cox2003topological, weaire2007instabilities, fortes2007instabilities}. 

If we move to $3$D, the problem is minimizing the area functional and few exact result exist. The only rigorous one is the proof of {\em the Double Bubble Conjecture} \citep{hutchings2002proof}, which states that the standard double bubble provides the least-area enclose and separates two regions of prescribed volume in $\R^3$. As regarding the numerical results, Kelvin \citep{thomson1887lxiii} proposed an optimal candidate structure with {\em identical cells}, which has been numerically refuted in \cite{weaire1994counter}: by numerical calculations, an agglomerate of two different types of bubbles has less perimeter, for fixed area.

The aim of this work is to study the symmetry break of a eight bubble compaction in $3$D, {\em i.e.} we want to investigate the mechanical cues driving the three-dimensional packing of $8$ bubbles displayed in a recalling foam structure. To circumvent the difficulty of a variational approach, {\em i.e.} the minimization of an area functional satisfying some geometrical constraints, we follow a different strategy. First of all, in Section \ref{sec:geometry}, we fix the geometrical  arrangement of the eight bubbles as the solution of the Tammes' problem \citep{tammes1930origin}: we exploit a geometrical principle of maximal mutual distance between neighbour points on a spherical surface obtaining seven symmetrical peripheral spheres tangent to the central one \citep{melnyk1977extremal}. Then, compaction is produced by packing the outer bubbles along the radial direction of the aggregate. The obtained agglomerate recalls the foam structure \citep{cantat2013foams}: the central sphere is completely covered, while the peripheral ones have a free-curved surface. While our construction does not ensure that the obtained final configuration is the minimal one, we will prove that the our tessellation on the central sphere is the optimal one among all the possible \citep{brinkmann2007fast} according to physical assumptions: the liquid/liquid interface is favoured versus the liquid/gas one and it maximizes the volume \citep{weaire2001physics}.\\
Fixing the geometrical arrangement, in Section \ref{sec:mechanical_balance}, we look for the balance of forces that produces such a configuration \citep{bikerman1973formation}. Enforcing the mechanical equilibrium and the conservation of volume, we derive the tensional balance laws on this geometrically optimal packing.
Finally, in Section \ref{sec:conclusione}, we discuss the results and we add few concluding remarks.

\section{Geometrical principle}
\label{sec:geometry}

\subsection{Spatial arrangement}

In this section, we introduce a geometrical principle that we exploit to describe the spatial arrangement 
of the $8$-bubbles configuration, to determine the position of the seven bubbles
surrounding the central one.  The coordinates of the peripheral bubbles centroids are given as solution of 
the classical {\em Tammes's Problem} \citep{tammes1930origin}: determine the arrangement of $n$ points 
on the surface of a sphere maximizing the minimum distance between nearest points ({\em maxmin principle}).
This is equivalent to determine (up to rigid rotations) the $n$ unit vectors $\{\r_i\}$  such that  
\begin{equation}
    \label{1}
    \lim_{m\rightarrow + \infty} \left \{ \frac{1}{|\r_i-\r_j|^m} \, : \, 1 \leq i < j \leq n \right \},
\end{equation}
is maximum, where the limit $m \rightarrow + \infty$ selects the distance among closest points only. \\ 
In our case $n = 7$; we want to find the position of seven points on a sphere 
with minimum distance from their nearest neighbours.
   
\begin{figure}[t!]
\begin{subfigure}{.5\linewidth}
		\centering
		\includegraphics[width=0.75\textwidth]{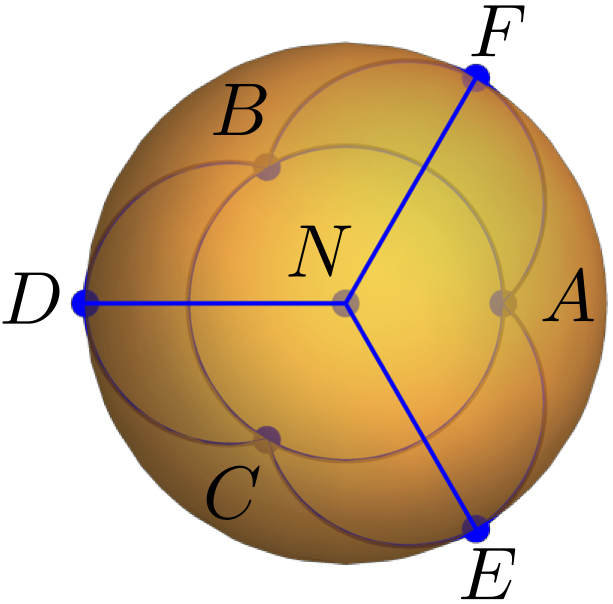}
		\caption{Top view}
		\label{fig:toptammes}
	\end{subfigure}%
	\begin{subfigure}{.5\linewidth}
		\centering
		\includegraphics[width=0.67\textwidth]{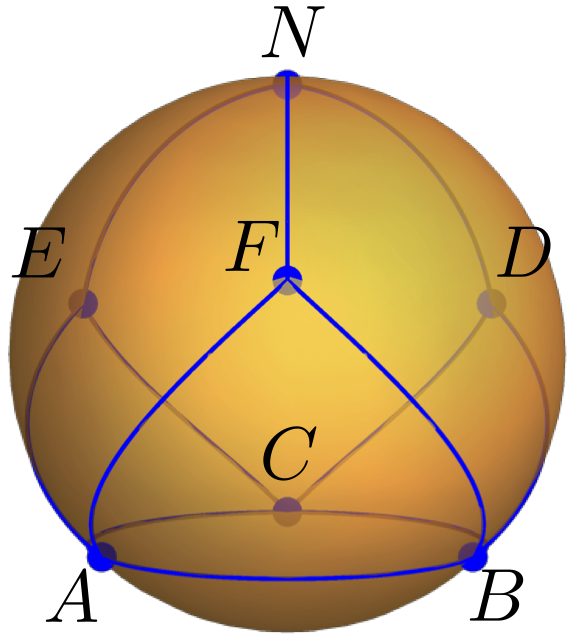}
		\caption{Side view}
		\label{fig:sidetammes}
	\end{subfigure}%
	\caption{(a) Top and (b) side view of the position and connections of the seven points on the
         spherical surface of the central bubble. The blue connections are the arcs of length $a$ 
	 defined by Tammes' construction.}
	\label{puntitammes}
\end{figure}

\begin{figure}
\begin{subfigure}{.5\linewidth}
		\centering
		\includegraphics[width=0.7\textwidth]{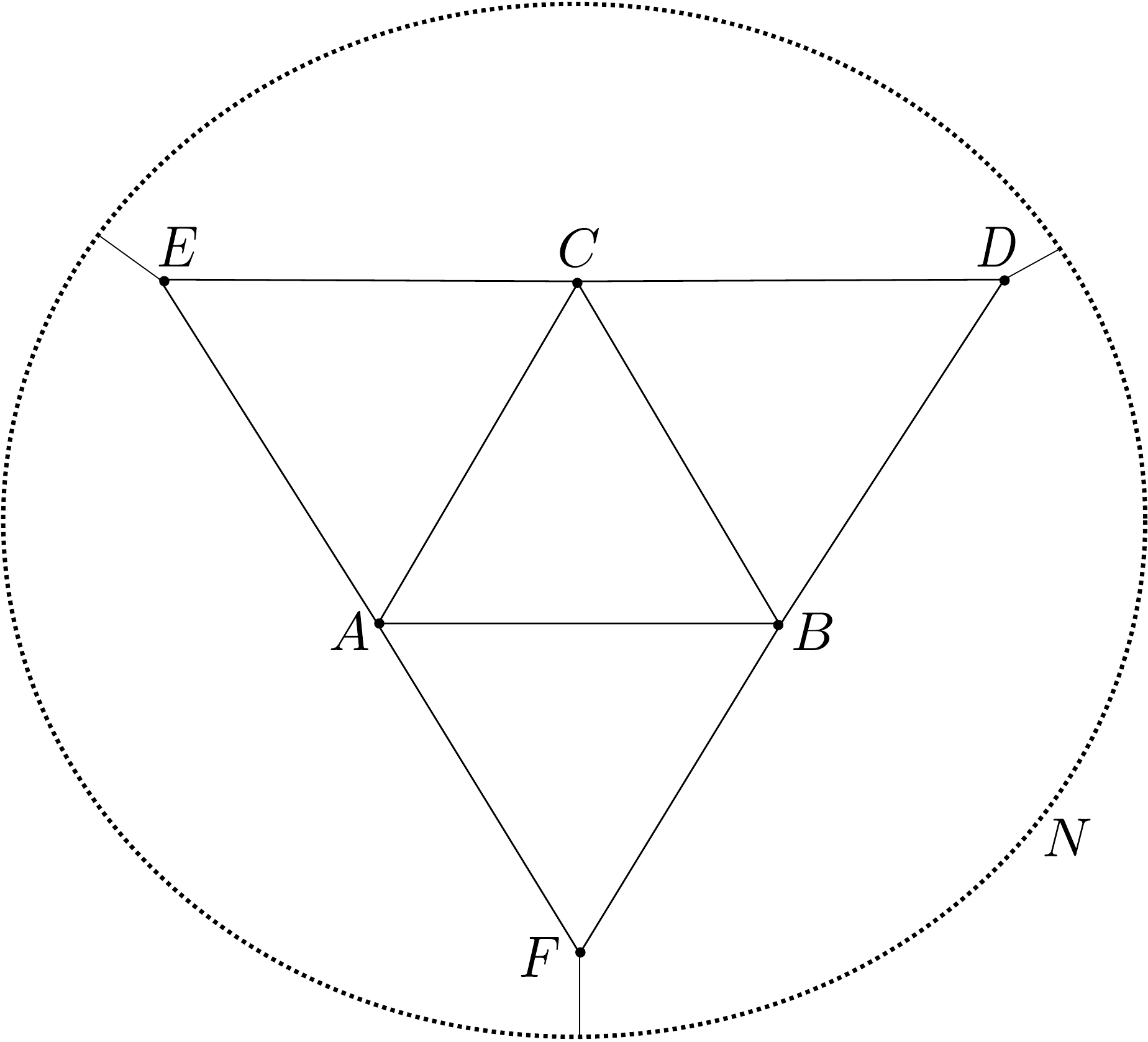}
		\caption{Stereographic projection}
		\label{fig:stereografica}
	\end{subfigure}%
	\begin{subfigure}{.5\linewidth}
		\centering
		\includegraphics[width=0.7\textwidth]{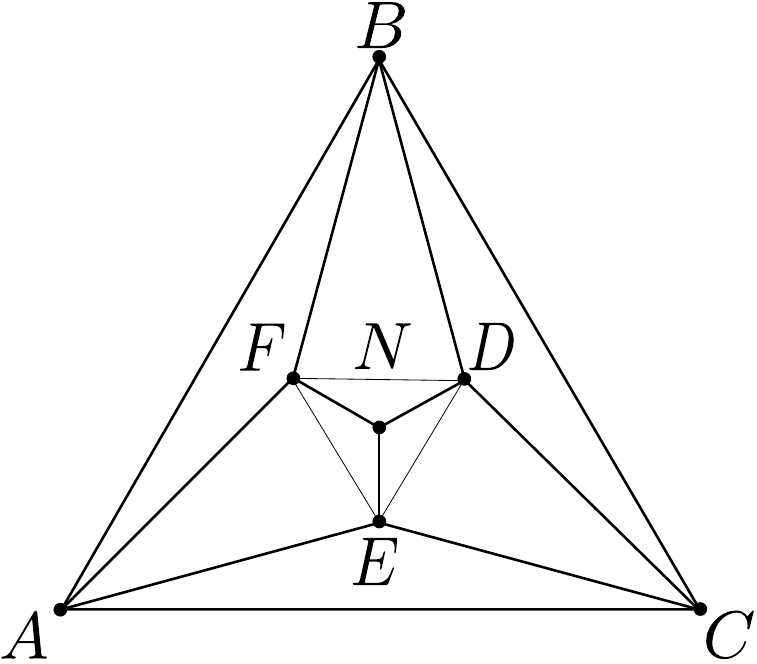}
		\caption{Spherical projection}
		\label{fig:sferica}
	\end{subfigure}
	\caption{Stereographic (a) and spherical (b) projections of the Tammes' points on the 
	spherical surface and their connecting arcs. Grey and thin lines, corresponding to arcs of length 1.34 $a$, 
	make the tessellation fully triangular.}
	\label{grafo}
\end{figure}

\begin{figure}
	\centering
	\includegraphics[width=0.45\textwidth]{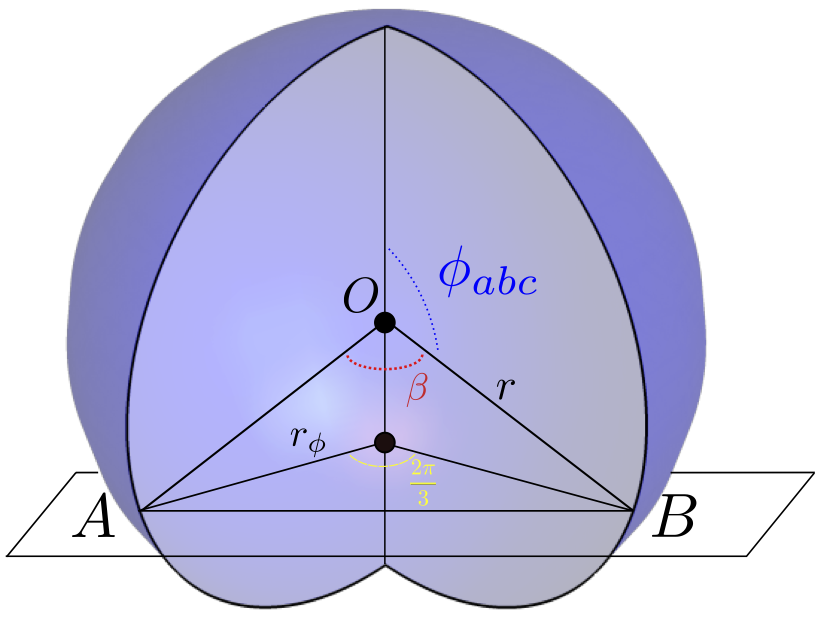}
	\caption{Geometrical sketch of the latitude of the points $A$, $B$ and $C$:           
	the central angle $\beta$, the radius $r$ of the sphere and the radius $r_{\phi}$ 
	of the circumference laying in the plane defined by the triplet of points.}
	\label{phiabc}
\end{figure}

Here, we exploit the {\em graph theory} \citep{west2001introduction} to find this maximizing configuration. 
A set of $n$ points on a sphere forms a graph $G$ of $n$ points connected by arcs of great circles 
of length $a$ \citep{schutte1951kugel}.
The maximal spatial arrangement of seven points can be obtained by the projective argument presented 
in \citep{schutte1951kugel}.  Consider a frame of reference centered in $O = (0,0,0)$ and the coordinates 
on the spherical surface $(r, \theta, \phi)$,  
where $\theta \in [0, 2\pi)$ (longitude) and $\phi \in [0,\pi]$ (latitude) on $S^2$, namely
$$
     S^2 = \{r, 0\leq \theta <  2\pi, 0\leq \phi \leq \pi \},
$$
where $r$ is the radius of the sphere $N$ and $S$ the North and South Pole, respectively. 
Three points $\left\{A,B,C\right\}$ are placed at the same latitude on the surface of the central sphere, 
such that they are connected by arcs of length $a$ and they form an equilateral triangle centered in $S$.  
Three more identical triangles are then created, adjacent to the former ones, with vertices $D,E$ and $F$: 
they share the same latitude too. The final step is then to connect $D,E$ and $F$ with $N$ and vary the radius
$r$ (for fixed $a$) until also the latter arcs have length $a$, \! \footnote{
	In a fully equivalent way, one can fix the radius $r$ and vary the chord length $a$.}
(see Fig. \ref{puntitammes}).
The associated extremal graph defines four triangles and three quadrangles on the spherical surface,  
as illustrated in Fig. \ref{grafo}. 
Fundamental relations of spherical trigonometry tell us that the internal angle of an equilateral 
spherical triangle is $\alpha = \frac{4\pi}{9}$, while the arc angle $\beta$ with respect to the centre 
of the sphere $\beta$ is given by \citep{berger2010geometry}
\begin{equation}
\label{beta}
         \cos{\beta} = \frac{\cos\frac{4 \pi}{9}}{1- \cos\frac{4 \pi}{9}},
\end{equation}
as illustrated in Fig. \ref{phiabc}.
This allows to find explicitly the linear
relation between arc length and radius, {\em i.e.} $a = \beta r$. \\
The length $\ell$ of the chord between closest points is 
$$
    \ell = 2 r \sin\left(\frac{\beta}{2}\right).
$$

\subsection{Coordinates}           
By construction, the spherical distance on $S^2$ of the points $D,E$ and $F$ from $N$ 
is equal to $a$, so their latitude  is the angle $\pqr = \beta$. 
For the triplet of points $\{A,B,C\}$, the calculations are a little bit more elaborated. Let $r_{\phi}$ be the 
radius of the circumference defined by the intersection of the sphere and the plane where $A,B$ and $C$ are.
The following relation holds
\begin{equation}
\label{5}
  \begin{aligned}
	  r_{\phi} = r \sin \left(\abc\right),
  \end{aligned} 
\end{equation}
where $\abc$ is the latitude of the points $A$, $B$ and $C$. \\
Since the chord length is the same for the spherical arc and for the in--plane circle, we find
\begin{equation}
\label{6}
               \frac{\ell}{2} = r_\phi \, \sin\left(\frac{1}{2}\frac{2 \pi}{3}\right) =   
               r \sin \left(\frac{\beta}{2} \right).
\end{equation}
By combining Eqs. \eqref{5} and \eqref{6} we get 
\begin{equation}
    \sin \abc = \frac{2}{\sqrt{3}} \sin \left(\frac{\beta}{2}\right).
\end{equation}
Summarizing, the coordinates of the seven Tammes points depicted in Fig. \ref{puntitammes} are
\begin{equation}
\begin{aligned}
\label{8}
   A &= (r,0,\abc) \qquad  B=\left(r, \frac{2\pi}{3},\abc \right) \qquad C = \left(r, \frac{4\pi}{3}, \abc \right),   \\
   D = (r, \frac{\pi}{3},\pqr)  & \qquad E = \left(r, \pi, \pqr \right) \qquad F=\left(r, \frac{5\pi}{3},\pqr \right) 
	\qquad N = (r,0,0).
\end{aligned}
\end{equation}
\begin{figure}
\begin{subfigure}{.5\linewidth}
		\centering
		\includegraphics[width=0.8\textwidth]{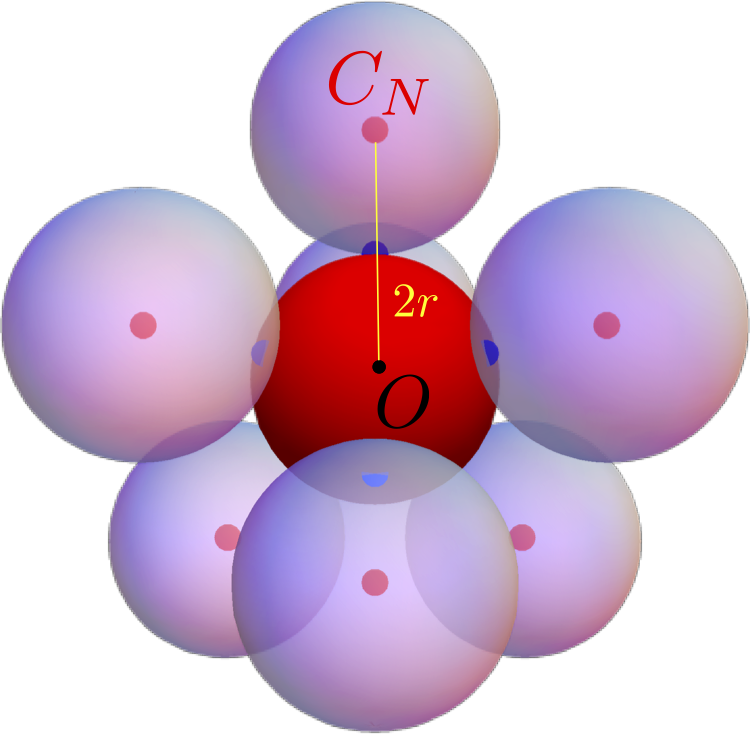}
		\caption{Before compaction}
		\label{fig:beforecompaction}
	\end{subfigure}%
	\begin{subfigure}{.5\linewidth}
		\centering
		\includegraphics[width=0.8\textwidth]{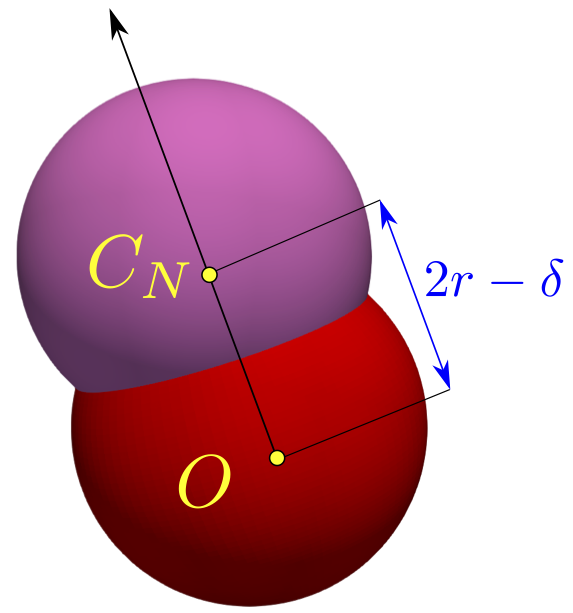}
		\caption{Compaction process}
		\label{fig:compactionprocess}
	\end{subfigure}
	\caption{(a) Initial arrangement of the bubbles before compaction. Tangent points are denoted by a blue dot, red dots denote the center of each sphere.
	 (b) Sketch of the ``compaction process'' between two bubbles driven by the parameter $\delta$.}
	\label{packing}
\end{figure} 
This configuration, given by the {\em maxmin principle} \eqref{1}, is here adopted as the ideal reference bubble 
arrangement:  seven spheres are tangent to the former one in the Tammes' points, 
as illustrated in Fig. \ref{fig:beforecompaction}.

\subsection{``bubble compaction'': tiling the central sphere} 
\begin{figure}[h!]
\begin{subfigure}{.5\linewidth}
		\centering
		\includegraphics[width=0.8\textwidth]{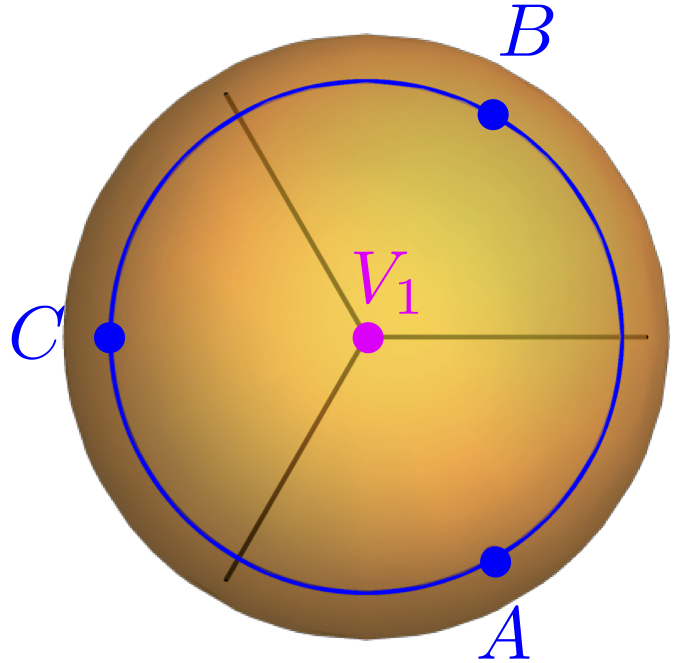}
		\caption{Bottom view}
		\label{fig:bottomtassellazione}
	\end{subfigure}%
	\begin{subfigure}{.5\linewidth}
		\centering
		\includegraphics[width=0.8\textwidth]{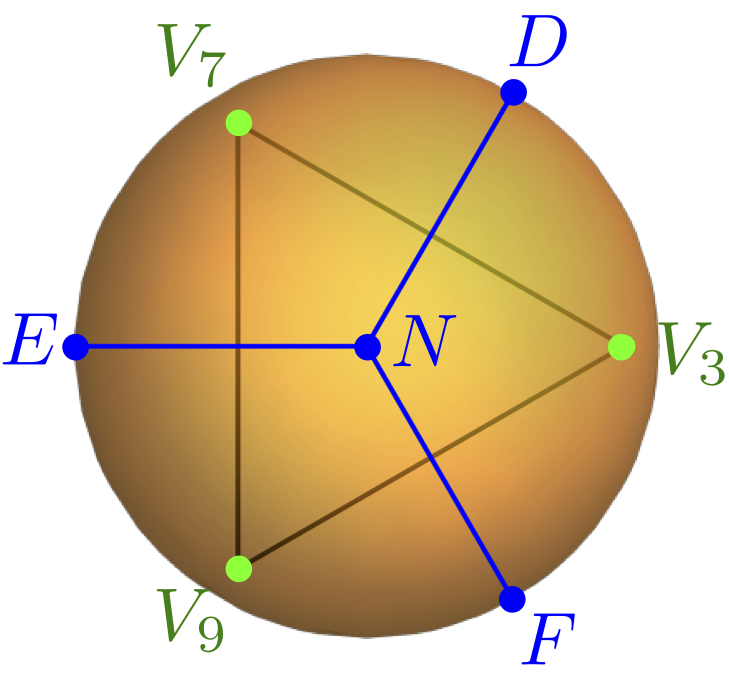}
		\caption{Top view}
		\label{fig:toptessellation}
	\end{subfigure}
	\caption{(a) Bottom view of the construction of the tessellation. (b) Top view of the construction of the tessellation.}
	\label{fig:costruiscolatassellazione}
\end{figure}

The tessellation of the spherical surface illustrated in the previous sections, is composed by four equilateral triangles and three quadrilaterals, see Fig. \ref{fig:stereografica}. However it can be made of triangles only by connecting points $\{D,E,F\}$, see Fig. \ref{fig:sferica}. The triangle $\{D,E,N\}$ is not equilateral, since the distance between $D$ and $E$ is equal to $1.34 \, a$.
On the basis of such a triangular tessellation we can produce a {\em dual} tessellation connecting 
the circumcenters of the triangles: the locus where the axis of the edges cross each other (Fig. \ref{fig:costruiscolatassellazione}).

The Tammes' points are the centroids of the polygons that define the dual tessellation 
(see Fig. \ref{fig:costruiscolatassellazione} and for more mathematical details Appendix \ref{a1}). \\ 
The bubble packing is obtained ideally moving each peripheral bubble, initially tangent to the central one 
in the Tammes' points, towards the origin $O$ along the radial direction, as illustrated 
in Fig. \ref{fig:compactionprocess}, while enforcing the volume conservation. In other words, to pack the 
bubbles aggregate we generate a collection of flat surfaces of contact among bubbles 
starting from the {\em maxmin} distribution of the tangent points: each peripheral bubble adheres
to the central one moving centripetally, see Fig. \ref{fig:compactionprocess}.  At the same time we  
shuffle the peripheral and the central spheres to preserve 
the initial volume $\V$. The contact surfaces between central and peripheral bubbles obtained by such
a {\em dive and shuffle} procedure are nothing but the polygons obtained connecting the points
of the dual tessellation defined above.
\begin{figure}[h!]
	\begin{subfigure}{.5\linewidth}
		\centering
		\includegraphics[width=0.78\textwidth]{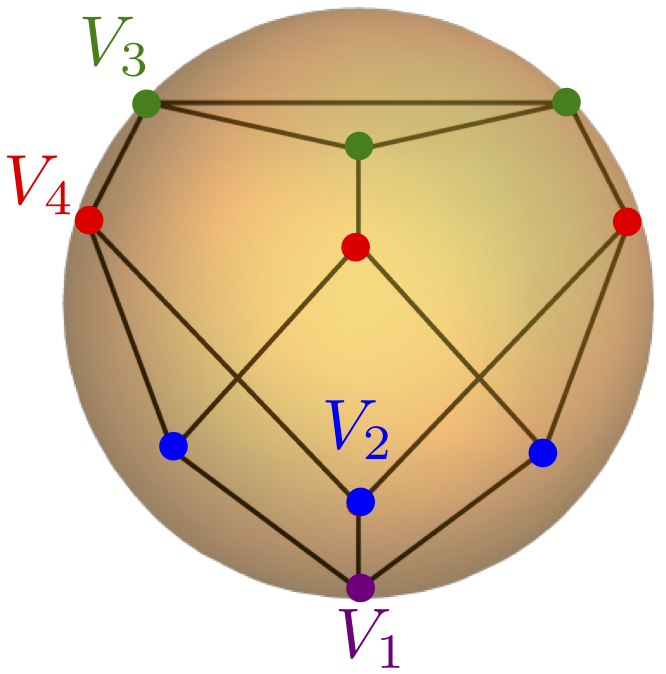}
		\caption{Independente nodes}
		\label{fig:nodiindependenti}
	\end{subfigure}%
\begin{subfigure}{.5\linewidth}
	\centering
	\includegraphics[width=0.8\textwidth]{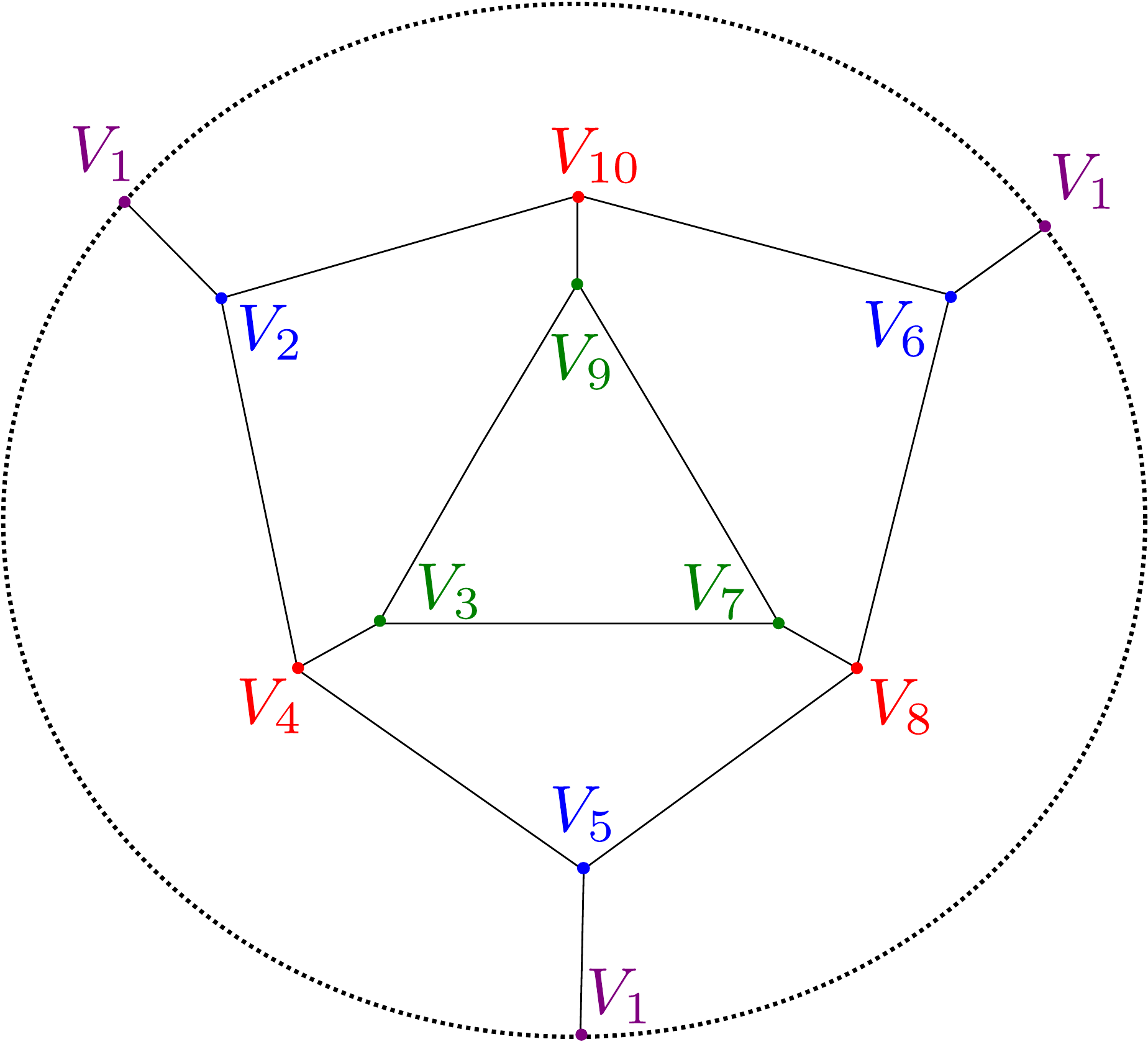}
	\caption{Stereographic projection}
	\label{fig:stereotassellazione}
\end{subfigure}
	\caption{(a) Sketch of four independent nodes $V_1$, $V_2$, $V_3$ and $V_4$ on the tessellation, 
		highlighting the corresponding symmetry group. (b) Stereographic projection of the dual tassellation.}
	\label{nodes}
\end{figure}

The final configuration is a tiling of the spherical surface of the central bubble with 
seven polygons, as illustrated in Fig. \ref{nodes}:
\begin{itemize}
\item an equilateral triangle centered in $N$ (area $\simeq 0.66 \,a^2$),
\item three quadrilaterals centered in $A,B,C$ (area $\simeq 0.7 \,a^2$), 
\item three pentagons centered in $D,E,F$ (area $\simeq 0.74\,a^2$).
\end{itemize}

\subsection{Geometrical optimality}
While the produced polygonal surface covers the central sphere, it naturally arises the question 
if such a tiling is optimal according to some suitable criterion. 
The problem to cover a spherical surface with polygons is old, rigorous results dating back to Cauchy 
\citep{cauchy1813recherches}.  In the present context, all the bubbles are identical 
and it is therefore tempting the idea to cover the central bubble with identical polygons. Unfortunately,
this is not allowed by Euler's Polyhedron Formula\footnote{Euler's Polyhedron Formula has been proved 
by Cauchy \citep{cauchy1813recherches} and it gives a relation among the number of faces, edges and 
vertices of a polyhedron, such as
$$
   F+V-E = 2.
$$
The number of vertices and edges are related with the number of faces faces $F$ as follows
$$
   E = \frac{Fn}{2} \qquad V = \frac{2E}{m} = \frac{Fn}{m},
$$
where $n$ is the number of edges of the polygon at hand, while $m$ is the number of faces 
which insist on the same vertex.  We are interested in the case $F=7$. By elementary calculations 
one can easily see that there is no $n$ for which a suitable integer $m$ exists.}.\\
There is no regular heptahedron. In order to prove if our tiling is optimal, we can construct all the convex polyhedra with $7$ faces which can be inscribed into a sphere of fixed radius $r$. By the software Plantri \citep{brinkmann2007fast}, we find $34$ convex polytopes with seven faces. They can be classified in terms of number of edges and number of vertices, see Table \ref{tab:nedges_nvertices}.
\begin{table}[h!]
	\centering
	\begin{tabular}{ |c| c|c|}
		\hline
		Number of vertices & Number of polyhedra & Number of edges  \\ \hline
		$6$&$2$ & $11$\\
		$7$& $8$& $12$\\
		$8$& $11$& $13$\\
		$9$& $8$& $14$\\
		$10$& $5$& $15$\\\hline
	\end{tabular}
	\caption{Classification of the $34$ convex polytopes with respect to the number of vertices (from left to right) or in terms of edges (from right to left).}\label{tab:nedges_nvertices}
\end{table}

The packing rearrangement of soap bubbles is dictated by both geometrical and mechanical motivations. Moreover, the optimal configuration is the one which minimizes the energy maximizing the volume. From the physics of the problem, we know that creating an interface liquid/liquid energetically costs less than one liquid/air \citep{weaire2001physics}. Hence, we can assume that the final shape of the central bubble has the maximum numbers of edges, {\em i.e.}
\begin{equation}
\label{eq:max_uno}
\widetilde{\mathcal{P}} = \max_{E=11}^{15}E_{\mathcal{P}_{i}} \quad i = 1,...,34.
\end{equation}
where $E_{\mathcal{P}_{i}}$ is the number of edges the $i$-polyhedron. In this way, we can reduce the number of polyhedra: we pass from $34$ convex polytopes with $7$ faces to just $5$ in which our tiling is included. In Appendix \ref{app:poliedri}, we show these $5$ polyhedra obtained and drawn by the software Plantri. So, we need another condition to select just one configuration. From the isoperimetric inequality, we know that the sphere is the solid that, fixing the area, it maximizes the volume and viceversa. For this reason, the shape of a single bubble is spherical. When two bubbles enter in contact, the surface of the agglomerate is lower than the surface of the two bubbles. 
Therefore, we search among the favorable energetically configurations, the one that has the maximum volume, such as
\begin{equation}
\label{eq:max_due}
\overline{\mathcal{P}} = \max_{i=1}^{5} \left(\mathcal{L}^3(\widetilde{\mathcal{P}_i})\right),
\end{equation}
where $\mathcal{L}^3$ is the volume measure. By numerically computing the five volumes, we find that the tiling obtained as the dual of the Tammes' one is the one with the maximum volume, {\em i.e.} the optimal one according to our criteria.
 \begin{rem}
 Since each polyhedron is not regular, we do not have an explicit formula to compute the volume.
 However, each polyhedron can be divided into $7$ pyramids, where the basis is a face. Using this geometrical argument, the total volume can be computed as the sum of the volumes of the pyramids.
 \end{rem}

\subsection{Surfaces, edges and vertices}
The dual tessellation defines ten nodes and it belongs
to $\mathcal{C}_{3z}(1,3,3,3)$.  
\footnote{$\mathcal{C}_{nz}$ is the group 
     of a cyclic symmetry after a rotation $2\pi / n$ with respect to the axis $z$ 
     \citep{clare1986closest,sands1993introduction}: the configuration is invariant for rotations 
     of an angle $ 2 \pi/3$ around the $z$ axis. 
     The notation $(1,3,3,3)$ denotes how many nodes of the tessellation share the same longitude.} 
We denote the nodes of the dual tessellation on the basis of the vertices of the Tammes' triangles they
belong to, such as
$$
\begin{aligned}
        &V_1=(A,B,C), \\
	&V_2=(A,F,B), \quad V_3=(E,F,N), \quad V_4=(A,E,F), \\
	&V_5=(A,E,C), \quad V_6=(B,D,C), \quad V_7 = (D,E,N),\\
	&V_8= (B,D,E), \quad V_9= (D,F,N), \quad V_{10}=(B,D,F).
\end{aligned}$$  
Because of the symmetry of the problem, there are only four independent nodes, as depicted in Fig. \ref{fig:nodiindependenti}.

The dual tessellation is formed by 15 edges, only 4 of them being independent. 
Each edge is identified by the polygons it belongs to on the tiled surface and 
the boldface denotes the unit vector parallel to the edge. 
Therefore, $\c_1=\overline{V_1V_2}$ denotes the edge between 
two quadrilaterals, $\c_2=\overline{V_2V_3}$ separates a quadrilateral and a pentagon, 
$\c_3=\overline{V_3V_4}$ separates a pentagon and a pentagon and $\c_4=\overline{V_2V_3}$ 
is between a pentagon and the triangle. \\ 
We generate a three dimensional structure projecting radially the dual tessellation, by an height to be 
fixed later on the basis of volume conservation arguments. 
Each vertex of the tessellation on the central bubble has therefore a corresponding outer one
that we denote by $V_i^h$, $i=1,2,3,4$.  
The connection between inner, outer and side surfaces is defined by two classes of 
edges:
\begin{itemize}
	\item $\c_5 = \overline{V_1V_1^h}$, $\c_6 = \overline{V_2V_2^h}$, $\c_7 = \overline{V_3V_3^h}$, 
	$\c_8 = \overline{V_4V_4^h}$ point radially, 
	\item $\c_9 = \overline{V_1^hV_2^h}$, $\c_{10} = \overline{V_2^hV_3^h}$, $\c_{11} = \overline{V_3^hV_4^h}$, 
         $\c_{12} = \overline{V_2^hV_3^h}$, are parallel to the ones on the tessellation of the central bubble.
\end{itemize}
At this stage the geometrical characterization of the 8-bubbles configuration derived on the basis of a 
maximum-minimum distance of the centroids of the peripheral bubbles is completed. The inner bubble 
has no free surface: it is surrounded by contact interfaces with other bubbles only. 
The external ones have the shape of a pyramidal frustum covered by a laterally cut spherical cap: 
the lower basis is the polygon generated by the adhesion with the central bubble, lateral sides are flat too, 
their edges being radially oriented, the upper basis of the frustum is a radial projection of the lower one. 
The upper geometrical structure is a spherical vault on a polygonal frustum, 
intriguingly known since the Middle Age in Sicilian architecture \citep{garofalo2015absidi}.
The radius of the spherical cap and the height of the frustum are to specified on the basis 
of balance and conservation arguments discussed below.                   

\section{Mechanical balance}
\label{sec:mechanical_balance}
In this section, we compute the surface tensions that make the geometrical packing mechanically 
equilibrated.  We remark that the central sphere has only bubble--bubble contact interfaces, 
while the peripheral ones also possess a traction--free surface.  Each bubble--bubble interface and 
each free surface is characterized by a tension $\tau_i$, defined as the energy density per unit area of the liquid/liquid or liquid/air interfaces \citep{roman2010elasto}. Thus, we have ten unknown independent tensions $\tau_i$: three on the 
central bubble, four on lateral bubble-bubble interfaces and three at the free surface denoted by
\begin{equation}
\label{tensioni}
\tau_{Q}, \tau_{P}, \tau_{T},
\tau_{QQ}, \tau_{PQ}, \tau_{PP},\tau_{PT},
\tau^s_{Q}, \tau^s_{P}, \tau^s_{T},
\end{equation}
where the subscript identifies the surface of the polygon it applies to and 
the superscript $s$ specifies the tensions at the free surfaces.

\subsection{Tensional balance}
\begin{figure}
\begin{subfigure}{.5\linewidth}
		\centering
		\includegraphics[width=0.8\textwidth]{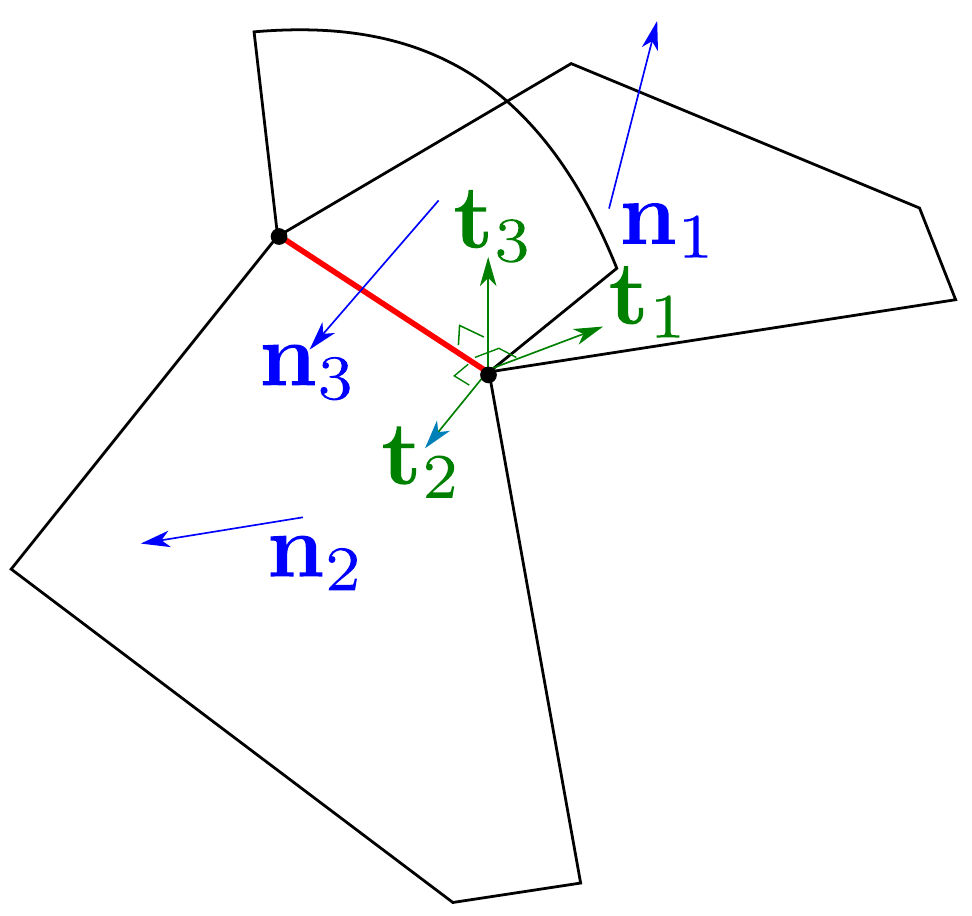}
		\caption{Balance on a edge}
		\label{fig:balanceedge}
	\end{subfigure}%
	\begin{subfigure}{.5\linewidth}
		\centering
		\includegraphics[width=0.8\textwidth]{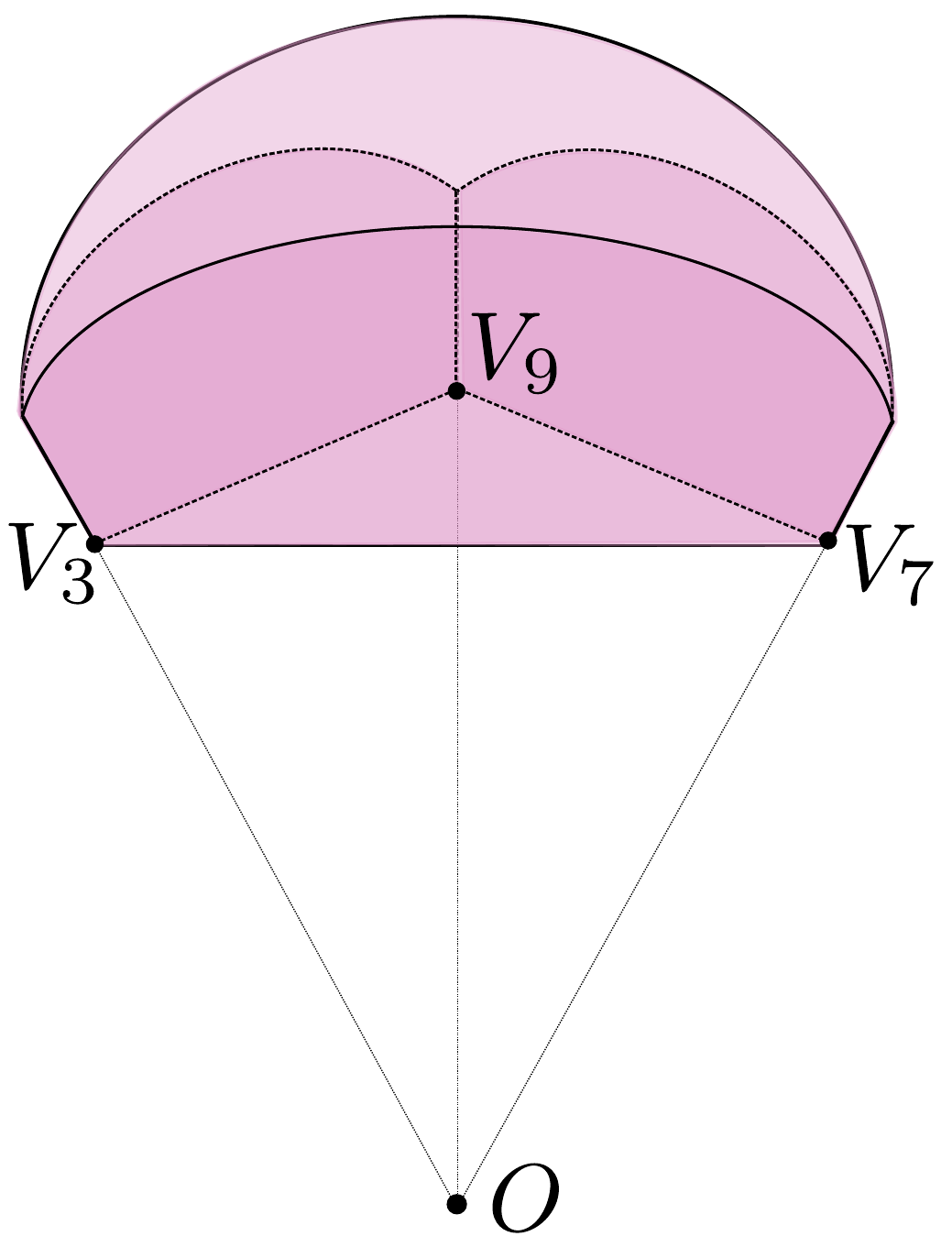}
		\caption{Volume of a peripheral sphere}
		\label{fig:volumelaterale}
	\end{subfigure}
	\caption{(a) Balance of tensions on an edge. (b) The bubble volume is the sum of a 
	pyramidal frustum (with pink side boundaries) plus the polygonal--basis vault standing 
	on it (light pink).
	}
\label{tensionvolume}
\end{figure}

First, we enforce the mechanical equilibrium imposing that the surface tensions are balanced 
on each independent edge $\c_i$, where $i=1,...,12$ (see Fig. \ref{fig:balanceedge}). 
Three (flat or curved) surfaces are attached to each edge, their local orientation 
being denoted by the normal unit vectors $\n^i_j$, $j=1,2,3$. 
The balance of tensions on each edge is defined by the sum of the tensions, oriented 
orthogonally to the edge and in-plane with the corresponding interface (see Fig. \ref{fig:balanceedge}). 
Therefore it must hold 
\begin{equation}
	\sum_{j=1}^3  {\mathbf t}_j \tau_j = 
	\sum_{j=1}^3 \c_i \times \n^i_j \tau_j =  \c_i \times \sum_{j=1}^3 \n^i_j \tau_j = 0 
	\qquad \Rightarrow \qquad \sum_{j=1}^3 \n^i_j \tau_j =0 \quad i=1,....,12 
\label{edge}
\end{equation}
Eq. \eqref{edge} defines 36 scalar equations, 12 of them being trivially null because all the
summed vectors are in the plane orthogonal to the edge under consideration. With the help of a symbolic software \footnote{We used Mathematica (Wolfram Inc., Version 12).}, we eventually find 
that, given the unit vectors $\n^i_j$ only 10 of them are independent.\footnote{
	The orientation of the normal unit vectors is not defined according any specific rule because 
	it is expected to affect only the sign of tension, that we know to be positive.} 
The equations are detailed in Appendix \ref{a2} in Tables \ref{sotto} - \ref{radiale} - \ref{sopra}.

The linear system Eq. \eqref{edge} is however not closed because, while the direction orthogonal 
to the flat surfaces is uniquely defined,  the edge contribution of the tension defined 
on the free surface depends on the curvature of the surface itself.  Curvature, tension 
and pressure gap on the free surface of the peripheral bubbles obey the Young-Laplace equation 
\citep{batchelor2000introduction} in the following way
\begin{equation}
\label{laplace}
           \Delta p = \frac{4 \tau^s_i}{R_i} \quad i = P, Q, T,
\end{equation}
where $R_i$ is the radius of curvature of the free surface of the $i$-th bubble and $\Delta p$ 
is the difference between the outer and the inner of pressure and there is an extra factor $2$ since the surface of the bubble is composed by two leaflets.  Since we have only three 
independent types of polygons on the tessellation, Eq. \eqref{laplace} gives three independent equations. \\

\subsection{Volume conservation}
\begin{figure}[t]
	\begin{subfigure}{.3\linewidth}
		\centering
		\includegraphics[width=1\textwidth]{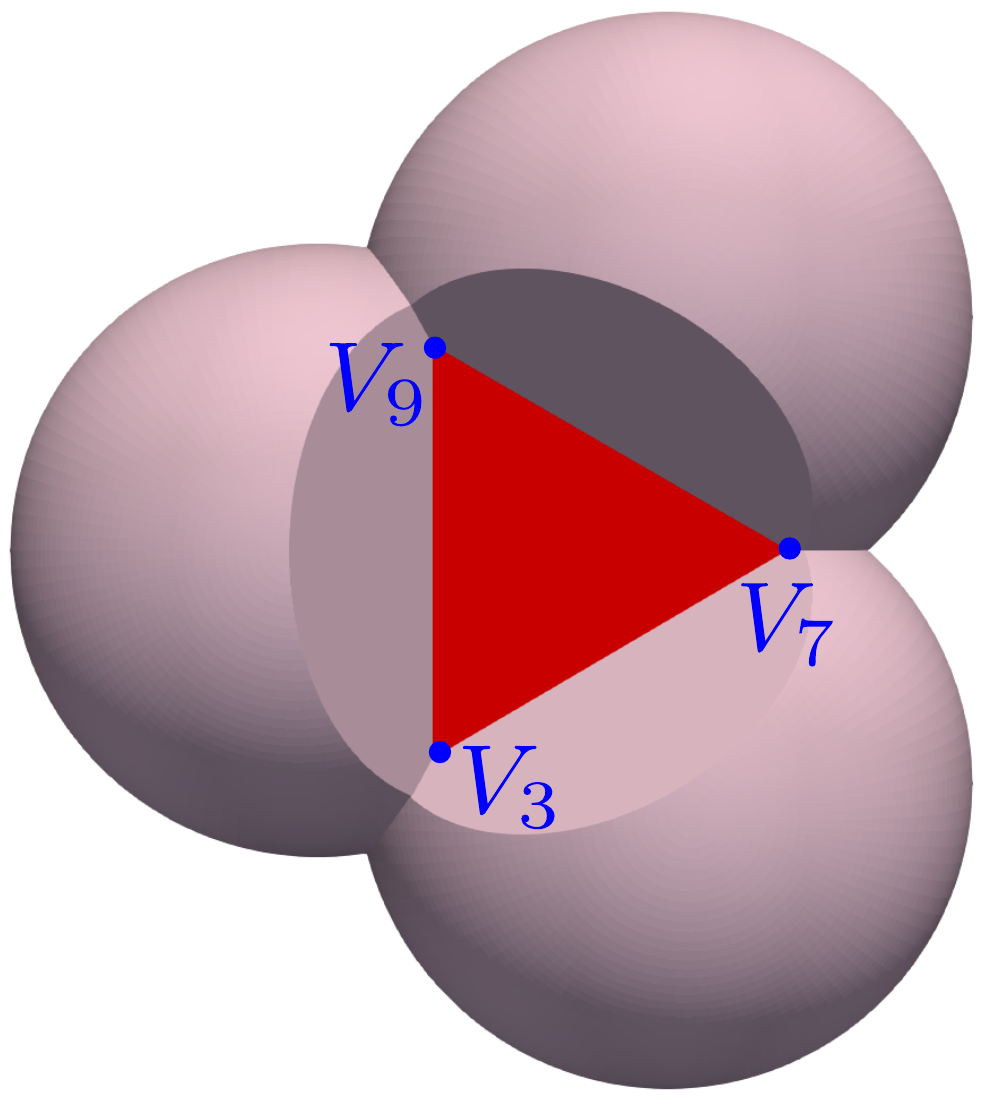}
		\caption{Triangle}
		\label{fig:triangle}
	\end{subfigure}%
	\begin{subfigure}{.3\linewidth}
		\centering
		\includegraphics[width=1\textwidth]{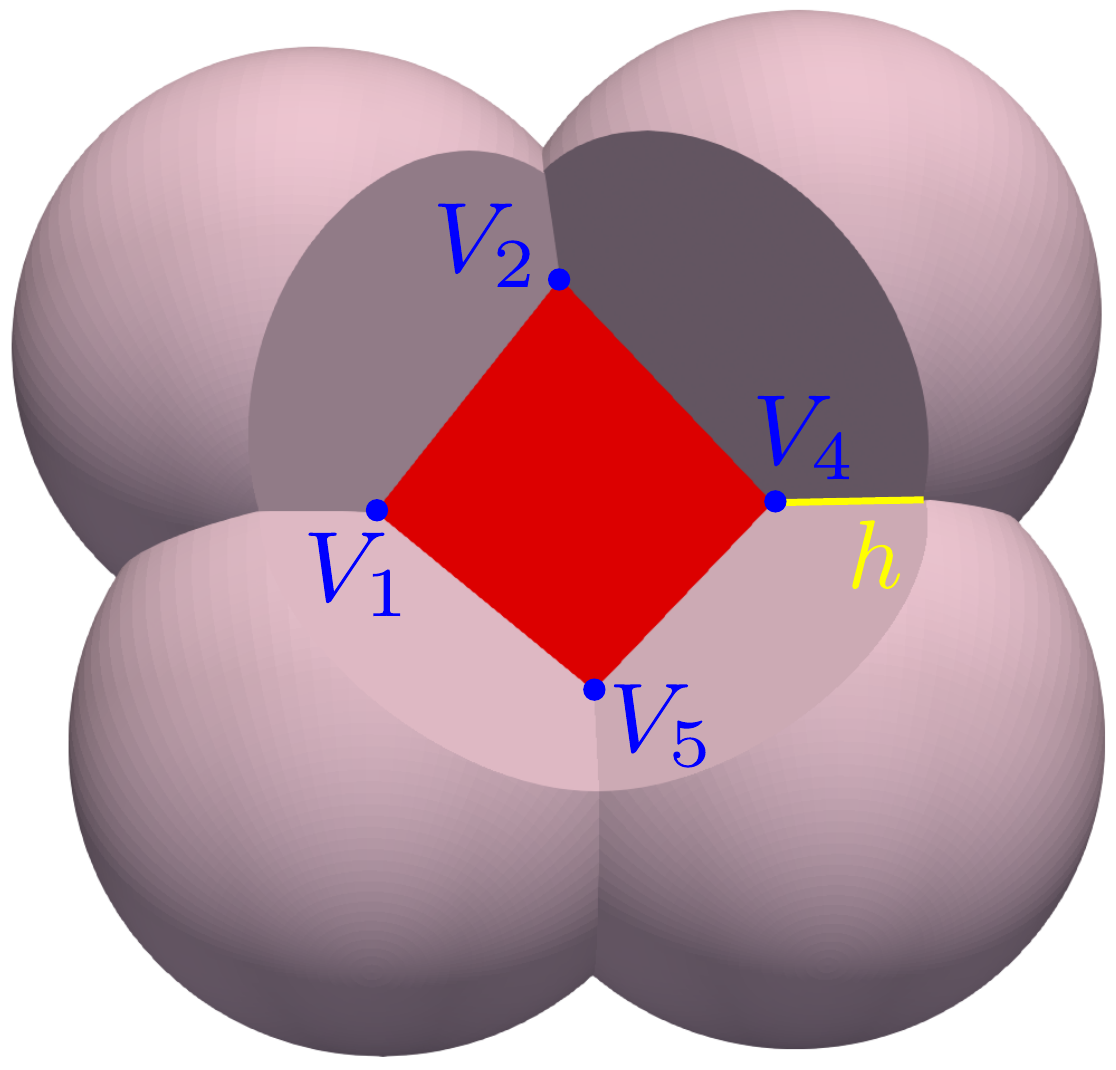}
		\caption{Quadrilateral}
		\label{fig:quadrilateral}
	\end{subfigure}
	\begin{subfigure}{.3\linewidth}
		\centering
		\includegraphics[width=1\textwidth]{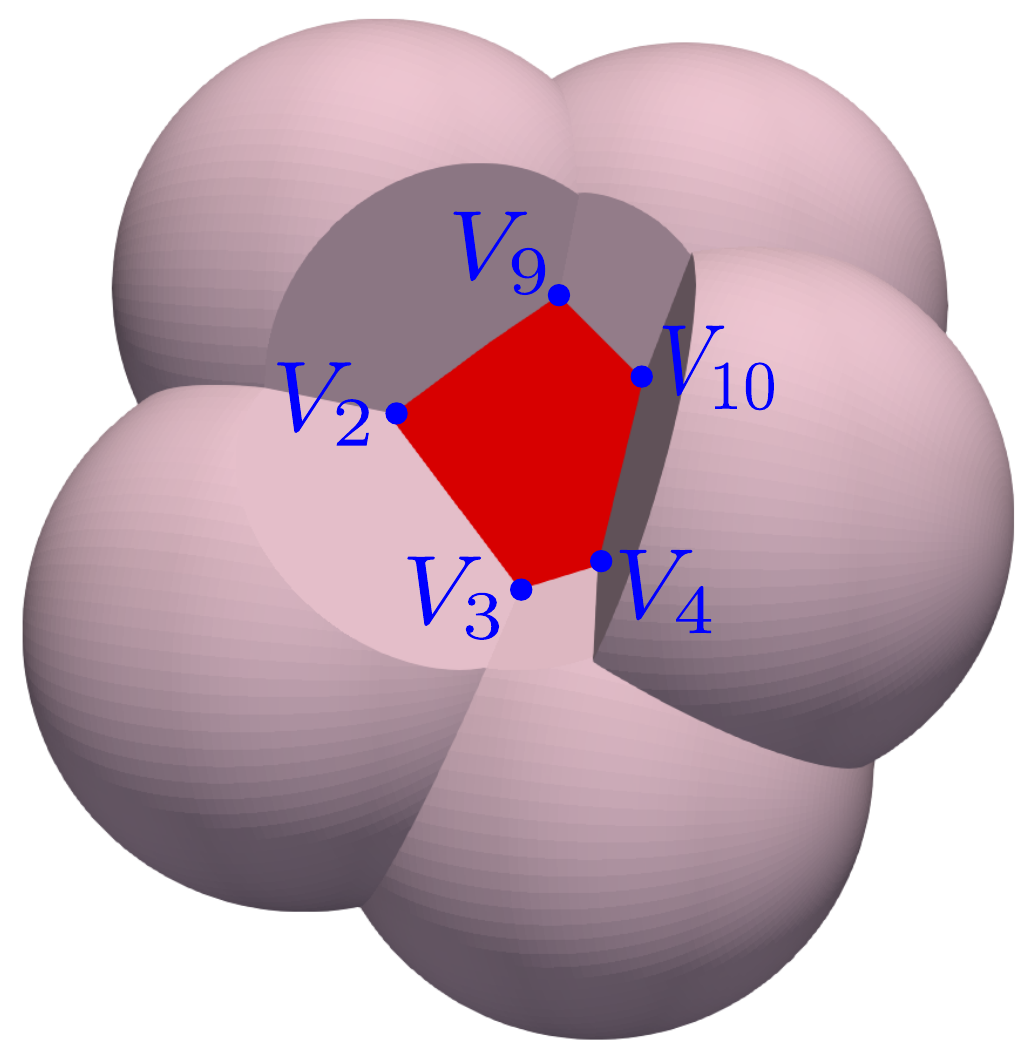}
		\caption{Pentagon}
		\label{fig:pentagon}
	\end{subfigure}
	\caption{Compactification of peripheral bubbles around (a) the triangle, (b) a quadrilateral 
		and (c) a pentagon.  The corresponding peripheral bubble is removed for the sake
		of graphical representation. The yellow segment represents the radial height $h$ of the intersection surface among three adjacent bubbles}
	\label{topbottomside tessellation}
\end{figure}

Finally, we have to impose the conservation of bubble volume under compaction.
While the central (packed) bubble is bounded by flat interfaces, the peripheral ones have 
the shape of a pyramidal frustum covered with a spherical vault (see Fig. \ref{fig:volumelaterale}). 
The basis of the pyramidal frustum are
\begin{itemize}
\item the interface with the central bubble;
\item its radially directed homothetic projection, by a factor $\frac{r+h}{r}$, where $h$ is the radial height of the intersection surface among three adjacent cells (see the yellow segment in Fig. \ref{topbottomside tessellation}).
\end{itemize}
The value of $h$ has to be fixed on the basis of volume conservation arguments: 
the sum of the pyramidal and apsal volumes 
must be equal to the common volume of all bubbles.

Details about the calculation of these volumes are given in Appendix \ref{a3}. 
As the area of each polygonal basis and the curvature radius of the apse are different, the radial height $h$ of the cells is not actually the same; however differences are below $1\%$.

 \subsection{Results}
We can solve the system of $16$ equations given by Eqs. \eqref{edge}-\eqref{laplace}, constrained to
volume conservation, with respect the $16$ unknowns: 10 tensions, 3 curvature radii and 3 heights. 
Using some experimental data coming from the foam literature, \citep{cantat2013foams}, we can fix the pressure difference $\Delta p = 50 \, {\rm Pa}$ and the radius of the round bubble as $r = 1 \, {\rm mm}$, before the compaction process.
Numerical solution of the nonlinear system of equations predicts the following surface tensions
\begin{equation}
\label{resultdim}
\left \{ 
\begin{aligned}
	&{\tau}_P = 51 \, \frac{m{\rm N}}{\rm m} &{\tau}_T =  45\, \frac{m{\rm N}}{\rm m}
	\quad&{\tau}_{Q} = 47 \, \frac{m{\rm N}}{\rm m} \\
	&{\tau}_{PP}= 69 \, \frac{m{\rm N}}{\rm m} &{\tau}_{QQ} = 58 \, \frac{m{\rm N}}{\rm m}
	\quad &{\tau}_{PT} = 63 \, \frac{m{\rm N}}{\rm m}&{\tau}_{PQ} =  65\, \frac{m{\rm N}}{\rm m}\\
	&{\tau}^s_P = 41 \, \frac{m{\rm N}}{\rm m} &{\tau}^s_T =35\, \frac{m{\rm N}}{\rm m}
	\quad&{\tau}^s_{Q} = 39 \,\frac{m{\rm N}}{\rm m}.
\end{aligned}
\right.
\end{equation}
The computed radii of curvature are  
\begin{equation}
\label{resuldim2}
\begin{aligned}
      &{R}_{P}= 3.5 \, {\rm mm}, &{R}_{T} = 2.5 \, {\rm mm}, &&{R}_{Q} =  3.1 \, {\rm mm}.
\end{aligned}
\end{equation}
The obtained surface tensions are consistent with experimental results \citep{cantat2013foams}: to create a soap bubble, the surface tension has to be less than the one of water, which is $\tau_{\rm water} \simeq 73 \,m{\rm N}/{\rm m}$, otherwise the bubble cannot exist. 
The obtained field of forces Eq. \eqref{resultdim} is the one at the equilibrium. We immediately notice that there is an anisotropy in the distribution of the field of forces: surface tensions of bubble-bubble interfaces with normal oriented in the circumferential direction of bubbles  
aggregate (second line of Eq. \eqref{resultdim}) are larger than the ones with normal unit vector 
pointing radially out of the aggregate (first and third line of Eq. \eqref{resultdim}).

This result supports our conjecture:
 the anisotropy in the mechanical cues may be the cause of the symmetry break, {\em i.e.} there might be a preferential direction of the next topological instability \citep{cox2003topological}. Indeed, from experiments and numerical results, it is known that a similar aggregate, due to some physical involved parameters, can develop an asymmetry or a topological transition \citep{weaire2007instabilities}. The study of the stability of this configuration is out of this paper. This result wants just to show that the distribution of forces in the equilibrium configuration itself is not symmetric, hence we can state that any small perturbations can change the rearrangement of forces inside the system and can develop a topological transition which breaks the starting symmetrical structure.

\section{Final remarks}
\label{sec:conclusione}
In this work, we studied the symmetry break of a particular configuration of $8$ spheres, showing the anisotropy in the distribution of the field of forces of the equilibrium position, that might possible originate a topological transition and break the symmetric structure of the starting agglomerate \citep{cox2003topological, weaire2007instabilities}.
This result can be applied to a physical situation, {\em i.e.} the study of foam, since the selected rearrangement of spheres remembers the one of soap bubbles in a single module of the foam structure \citep{cantat2013foams}. 

We considered $7$ identical spheres symmetrically surrounding a central one: their initial position is dictated by the solution of the Tammes' problem \citep{tammes1930origin}. Neglecting any dynamical process, the final configuration is obtained by a compaction process which results into a full tiling of the central sphere. By introducing physical criteria of optimality dictated by the energy minimality, Eq. \eqref{eq:max_uno}, and by the volume maximality, Eq. \eqref{eq:max_due}, we proved that our polyhedra is the optimal one among all the $34$ convex polytopes inscribed into a sphere with radius $r$ \citep{brinkmann2007fast}, since due to Euler Polyhedra Formula no regular heptahedrons exist \citep{cauchy1813recherches}. 

Fixing this geometrical arrangement, we looked for the force balance that realizes such a configuration: we computed balance of forces on every edge, we forced the conservation of volume (by experimental evidences soap bubbles can be assumed to be incompressible \citep{exerowa1997foam}) and we imposed the Laplace law on the possibly curved free surface. We obtained a force field, Eq. \eqref{resultdim}, which fullfils an acceptable physical range \citep{cantat2013foams}, but it shows an anisotropy in its orientation, see second line in Eq. \eqref{resultdim}. This result suggests that a difference in tension, generated by a purely mechanical principle, might be crucial for next topological transitions and for the development of anisotropies \citep{weaire2007instabilities}. In this respect, we conjecture that the expulsion if a single bubble, dictated by any small perturbations, in the {\em flower cluster} in $2$D \citep{cox2003topological} can be replicated in higher dimensions, breaking the symmetrical structure. 

Future efforts will be to devoted to reproduce this system both in a laboratory and numerically to study the dynamical evolution of this agglomerate.

The main novelty of this work is the application mathematical method, to a particular context, {\em i.e.} the symmetry break of a $8$-bubble compaction. In general. the study of the geometrical rearrangement and the change of shape of a configuration by mechanical and geometrical considerations might introduce a new non-destructive approach to better understand different physical phenomena. For instance, it can be used to design new meta-materials, where it is fundamental to know a priori the balance of forces, or to study the mitosis of cells. Indeed, just by knowing their geometrical rearrangement at a fixed stage, we can determine if the distribution of forces has an anisotropy which can favour the duplication process along a particular direction. This means that our method can give an insight on a more detail comprehension on the mechanics of morphogenesis of a variety of tissues.

\section*{Acknowledgements}
The author thanks Davide Ambrosi, Pasquale Ciarletta, Alfredo Marzocchi and Maurizio Paolini for helpful suggestions and fruitful discussions.
This work has been partially supported by INdAM$-$GNFM.

\bibliographystyle{abbrvnat}
\bibliography{refs}

\begin{appendices}
	\section{Dual tessellation of the central bubble}
\label{a1}
\begin{figure}[h!]
\centering
\includegraphics[width=0.35\textwidth]{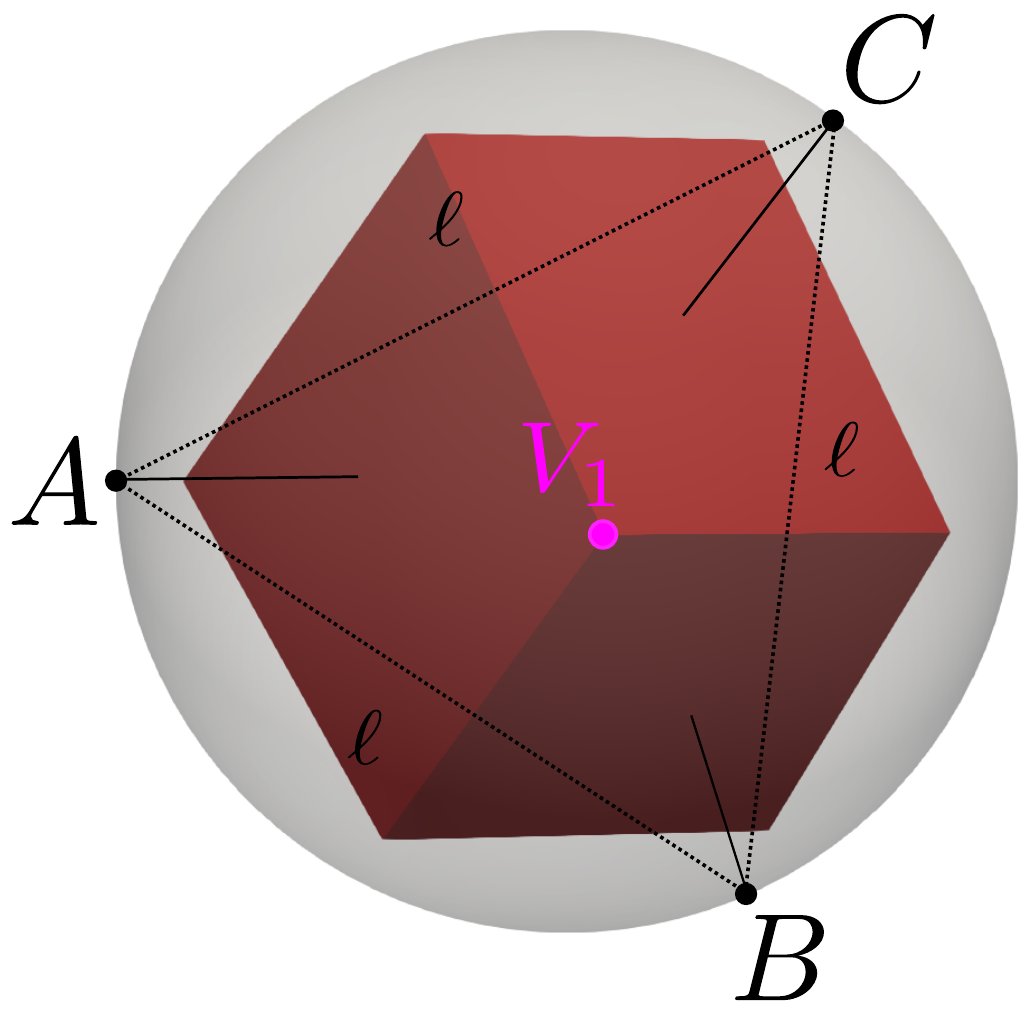}
\caption{Bottom view of the dual tessellation. The dashed lines represent the chord $\ell$ of the Tammes' 
	construction. The centroid of the triangle $A,B,C$ is the South Pole, called $V_1$ in the dual tessellation.}
\label{costruiscotassellazione}
\end{figure}

\begin{figure}[h!]
\begin{subfigure}{.3\linewidth}
		\centering
		\includegraphics[width=1\textwidth]{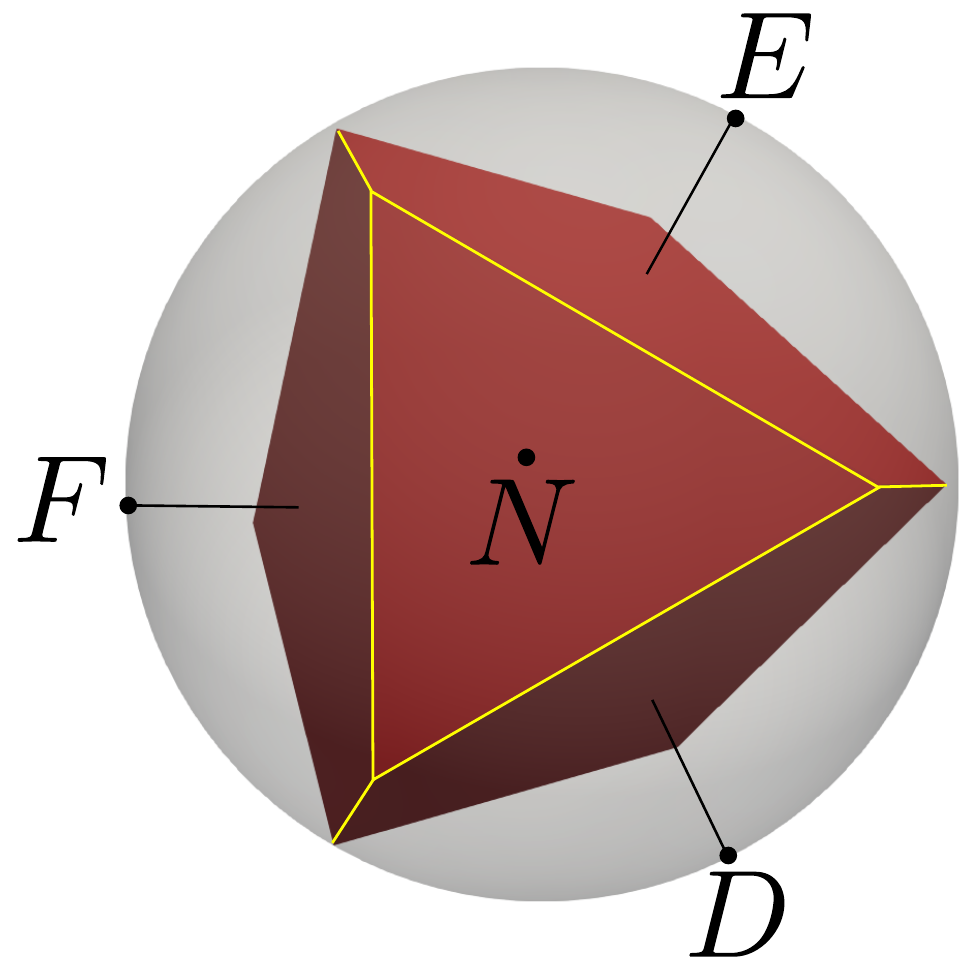}
		\caption{Top view}
		\label{fig:topcompatto}
	\end{subfigure}%
	\begin{subfigure}{.3\linewidth}
		\centering
		\includegraphics[width=1\textwidth]{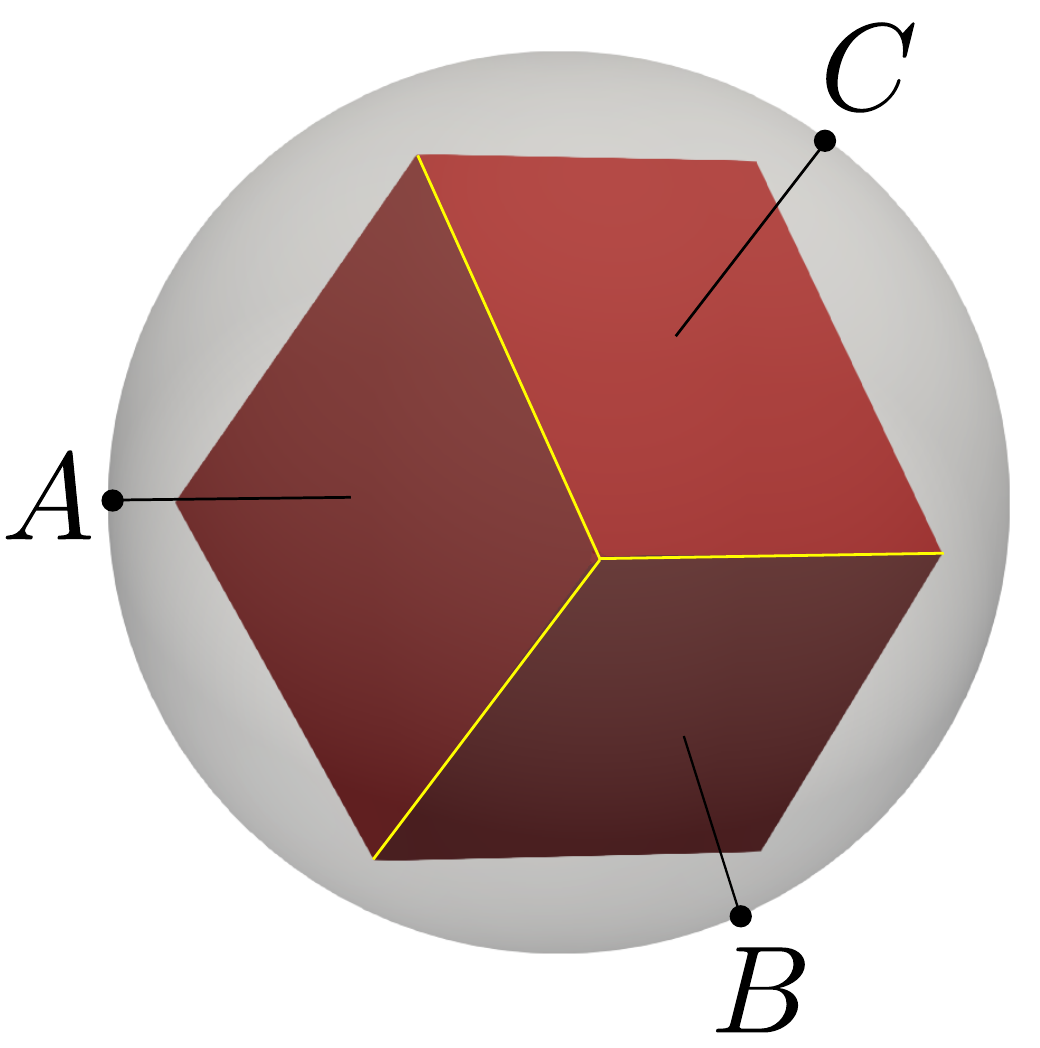}
		\caption{Bottom view}
		\label{fig:bottomcompatto}
	\end{subfigure}
	\begin{subfigure}{.3\linewidth}
		\centering
		\includegraphics[width=1\textwidth]{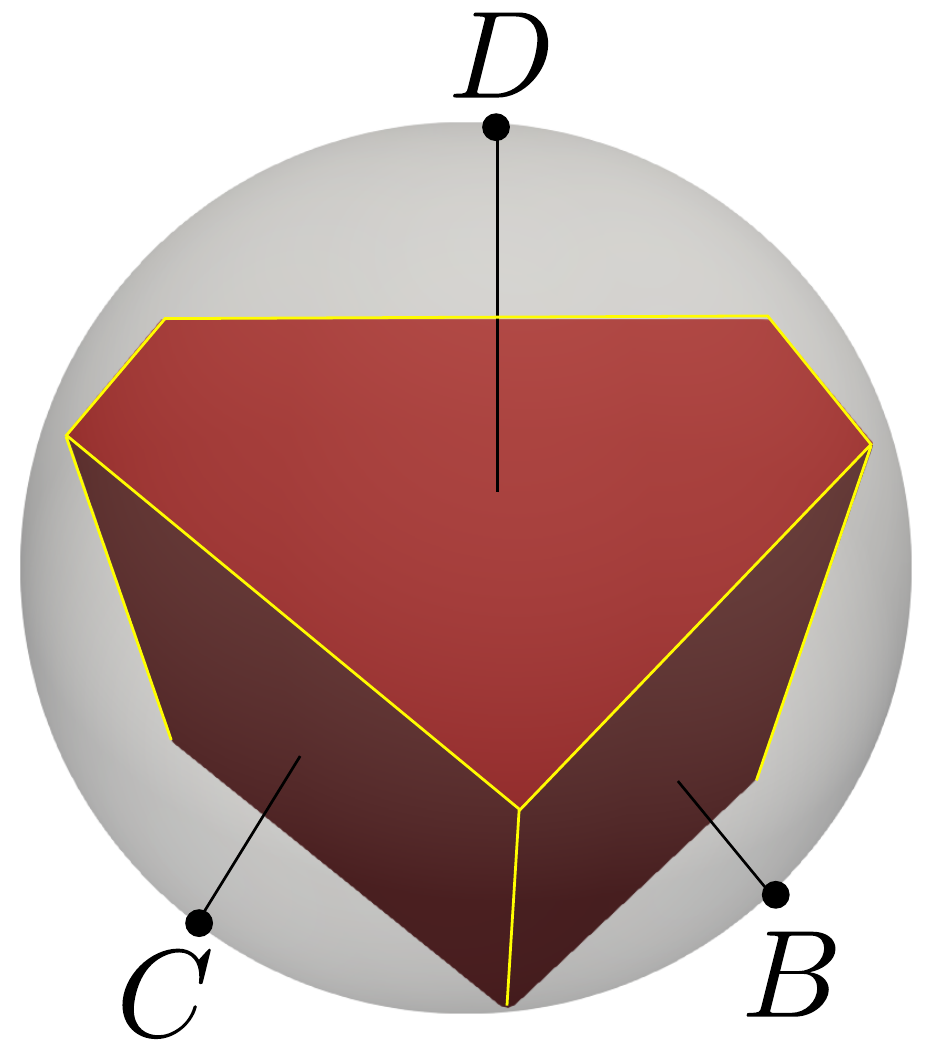}
		\caption{Side view}
		\label{fig:sidecompatto}
	\end{subfigure}
	\caption{Different views of the tessellation: (a) top, (b) bottom and (c) side view. 
	         Black dots indicate the Tammes' points. Irrespective of the graphic illusion,
		 the corners of the inscribed polyhedron are on the spherical surface; after 
		 restoring of the initial volume, they will be external. Yellow lines underline the sides of the different obtained polygons: (a) a triangle, (b) three quadrilaterals and (c) three pentagons.}
	\label{topbottomsidetessellation}
\end{figure}

A visual three-dimensional representation of the dual tessellation produced on the central sphere
by the bubble packing is depicted in Figs. \ref{costruiscotassellazione} and 
\ref{topbottomsidetessellation}. The polygons representing the flat bubble--bubble interfaces 
are here plotted inside the original spherical bubble of radius $r$ {\em before} reshuffling the polyhedron
to recover the original bubble volume. 

As we can see from Figs. \ref{costruiscotassellazione}-\ref{topbottomsidetessellation}, the projection of Tammes' points along the radial direction represents the centroid of each polygon on the dual tessellation. Viceversa, the vertices $V_i$ with $i = 1, \dots, 10$ are the centroids of the triangles of the modified Tammes' tessellation, see Fig. \ref{fig:sferica}.

 \section{Polyedra with seven faces}
 \label{app:poliedri}
\begin{figure}[ht!]
 	\begin{subfigure}{.3\linewidth}
 		\centering
 		\includegraphics[width=0.95\textwidth]{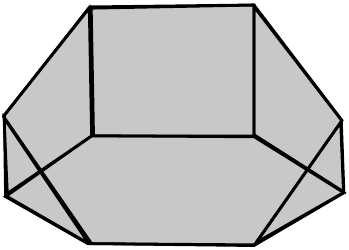}
 		\label{fig:ettaedro1}
 	\end{subfigure}%
 	\begin{subfigure}{.3\linewidth}
 		\centering
 		\includegraphics[width=1\textwidth]{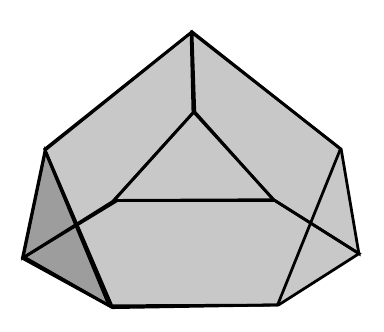}
 		\label{fig:ettaedro2}
 	\end{subfigure}
 	\begin{subfigure}{.3\linewidth}
 		\centering
 		\includegraphics[width=0.85\textwidth]{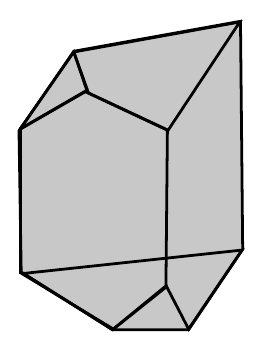}
 		\label{fig:ettaedro3}
 	\end{subfigure}\\
 	\begin{subfigure}{.5\linewidth}
 	\centering
 	\includegraphics[width=0.7\textwidth]{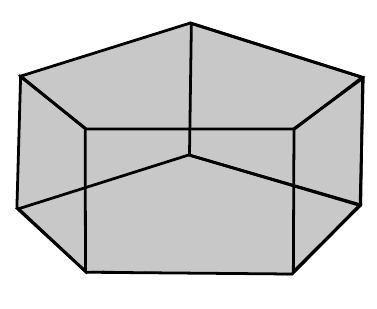}
 	\label{fig:ettaedro4}
 \end{subfigure}
	\begin{subfigure}{.5\linewidth}
	\centering
	\includegraphics[width=0.55\textwidth]{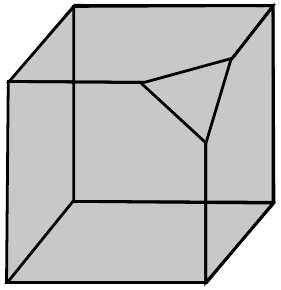}
	\label{fig:ettaedro5}
\end{subfigure}\\
 	\caption{Graphical representation, obtained through the software Plantri, of the five convex polygons among the $34$ inscribed into a sphere which satisfy the first optimal criterium Eq. \eqref{eq:max_uno}. The Tammes' dual tessellation is one of them, {\em i.e.} the last one.}
 	\label{fig:ettaedri_plantri}
 \end{figure}

 In this appendix, we want to show the five polyhedra among all the $34$ convex polytopes with seven faces inscribed into a sphere of a fixed radius $r$, which satisfies the first optimal criterium, {\em i.e.} Eq. \eqref{eq:max_uno}. By the software Plantri, we can draw them and their graphical representation is presented n Fig. \ref{fig:ettaedri_plantri}.
 
\section{Calculation of volumes}
\label{a3}
\begin{figure}[h]
	\centering
	\includegraphics[width=0.55\textwidth]{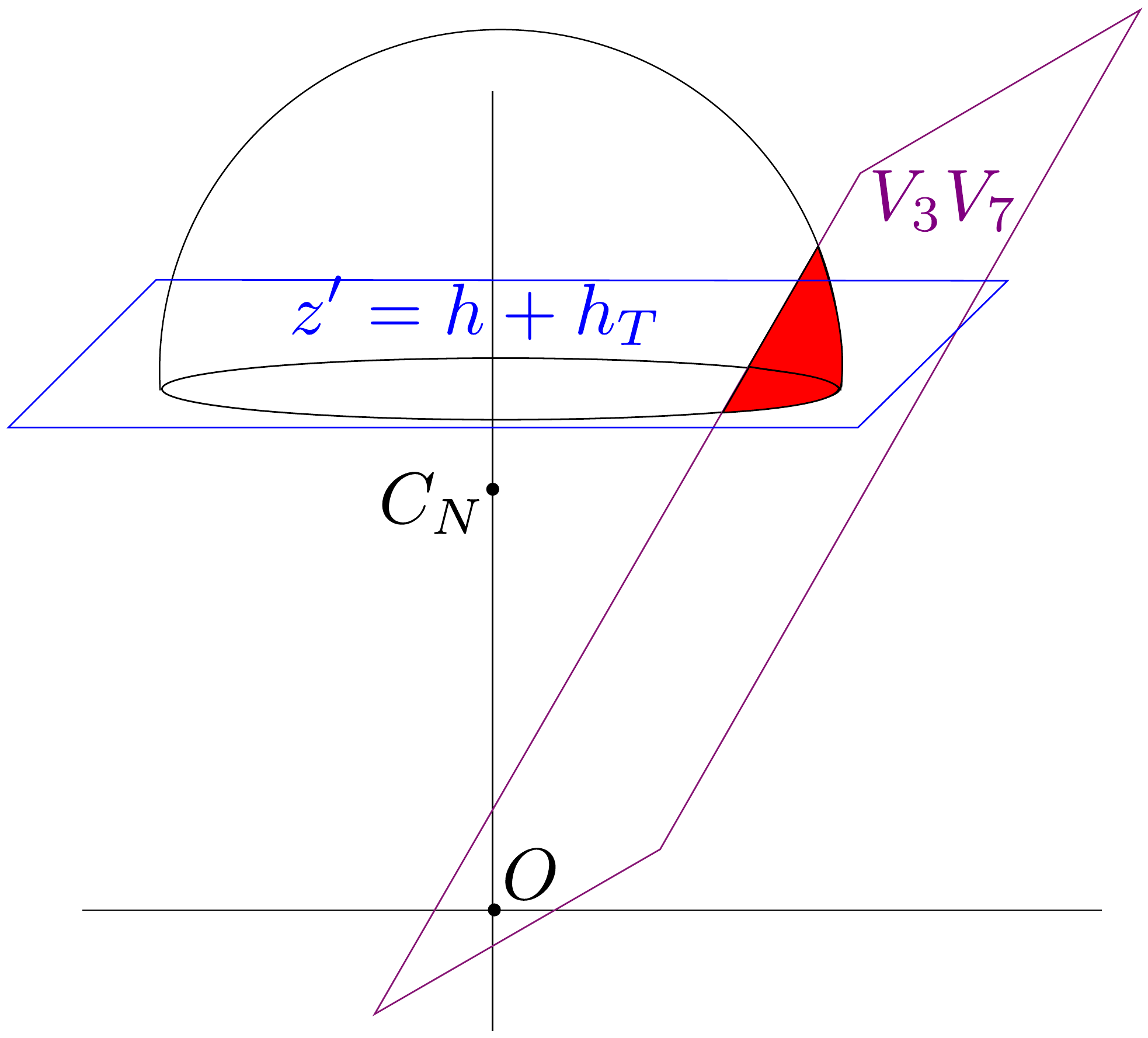}
	\caption{Geometrical representation of the lateral cut of the spherical cap.}
	\label{segmentocircolare}
\end{figure}

In order to compute the volume of the peripheral bubbles, we have to mathematical describe the shape of a pyramidal frustum covered by a vault. The vault is 
obtained radially cutting slices of a spherical cap drawn on the vertices of the external basis of the 
frustum, by prolongation of the lateral surfaces of the frustum itself 
(see Figs. \ref{fig:volumelaterale} and \ref{segmentocircolare}). From a mathematical point of view, this calculation is a bit technical since the final solid is not a known or a common one.\\

\begin{figure}
	\centering
	\includegraphics[width=0.55\textwidth]{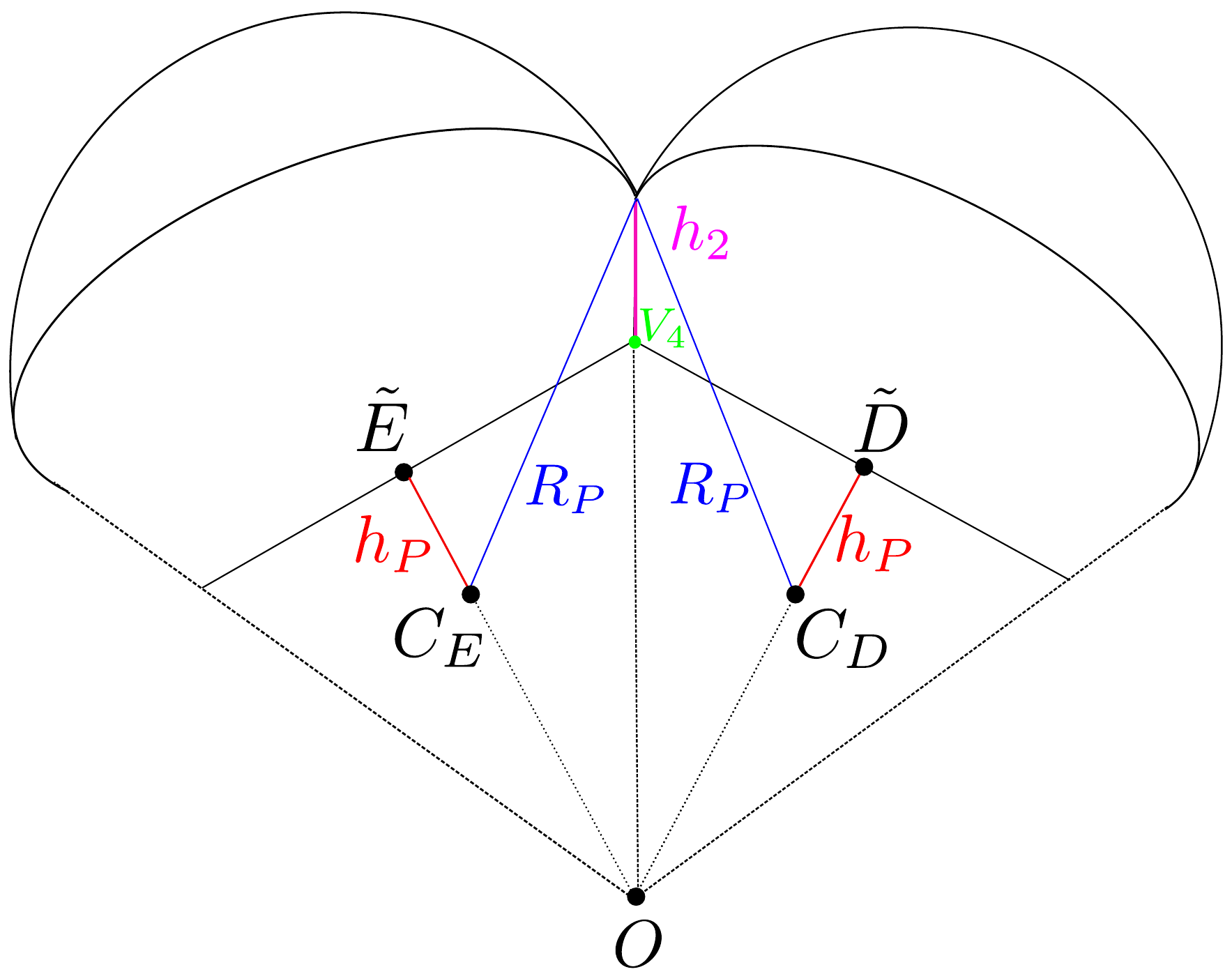}
	\caption{In the plane passing through the projected Tammes points $\tilde{D}$, $\tilde{E}$ and 
		the origin $O$, the points $C_{D}$ and $C_{E}$ are the centres 
		of the spheres of radius $R_P$, that eventually define the free surface of the bubble. 
		The side of the pyramidal frustum is $h_2$.}
	\label{palle1}
\end{figure}

First of all, we have to better clarify which are the involved unknowns. As it concerns the packing parameter, which is used to  define the position of the vertices on the free surface and it is obtained though the interaction among three surfaces, we a priori have four different values of packing parameter, {\em i.e.}
\begin{equation}
\label{hi}
h_1 = h(Q,Q,Q) \quad h_2 = h(P,P,T) \quad h_3 = h(Q,P,Q) \quad h_4 = h(Q,P,P),
\end{equation}
where the letters into brackets denote the interaction between three polygons, {\em i.e.} $P$ for pentagons, 
$Q$ for quadrilaterals and $T$ for the triangle.\\
We denote with the symbol $\,\tilde{}\,$ the centroids of the dual tessellation polygon: $\tilde A,\tilde B,\tilde C,\dots$,
the center of the sphere passing through the vertices of the homotetically projected polygons
is $C_i$, with $i = A,B,C,D,E,F,N$. Since we have three type of polygons, we make the calculations only on a representative 
of each class, such as on the quadrilateral $A$, on the pentagon $D$ and on the triangle $N$. \\
We define $h_P=\tilde D - C_D$, $h_Q=\tilde A- C_A$ and $h_T=\tilde N - C_N$  (see Fig. \ref{palle1}) 
the distance between the centre of the sphere associated with the vault curvature and the centroid 
of the polygonal inner basis of the frustum (the projection of the Tammes' points on the central bubble interfaces).
These distances are to be fixed in order to enforce 
conservation of volumes of the peripheral bubbles. \\ 
Depending on the type of adjacent polygons (pentagon-pentagon, pentagon-triangle,....), 
a different packing parameter $h_i$ with $i=1,2,3,4$ is expected (see Eq. \eqref{hi}). 
For illustrative purposes, here below we only show how the radial height $h_2$, obtained by the intersection among the sphere constructed on the triangle and on the two adjacent pentagons. We are going to show that it can be rewritten as a function of $h_P$  (Fig. \ref{palle1}). \\ 
The sphere of radius $R_P$, centered in $C_{D}$ is defined by the equation 
\begin{equation}
\label{S1}
(x-x_{C_{D}})^2+(y-y_{C_{D}})^2+(z-z_{C_{D}})^2=R_P^2,
\end{equation}	
where the coordinates of $C_{D}$ are
$$
C_{D} =\left(d\,x_{\tilde{D}},d\,y_{\tilde{D}},d\,z_{\tilde{D}}\right) 
\quad \hbox{with} \quad d= 1-\frac{h_P}{r}.
$$
Analogously, we construct the sphere on the adjacent pentagon $E$, so that
\begin{equation}
\label{S2}
(x-x_{C_{E}})^2+(y-y_{C_{E}})^2+(z-z_{C_{E}})^2=R_P^2.
\end{equation}
The same compation parameter $d$ scales the coordinates 
$\left(x_{C_{E}},y_{C_{E}},z_{C_{E}}\right)$ because both $D$ and $E$ are pentagons. \\ 
The intersection between the two spheres is a circumference, and the edge 
$\c_3 = \overline{V_3V_4}$ lies on it. In the same way, we can also consider 
another circumference obtained by the intersection of, for instance, the sphere centered in $C_{D}$ and the adjacent constructed on the triangle, {\em i.e.} centered in $C_{N}$.
The length of the side of the frustum $h_2$ is the distance between 
the node $V_4$ of the dual tessellation and the outer intersection point 
of the two circumferences defined above. 
The parametric representation of the radial line passing through 
$V_4 = (x_{V_{4}},y_{V_{4}},z_{V_{4}})$ is given by
\begin{equation}
\label{L1}
\left \{ 
\begin{aligned}
x = tx_{V_{4}},\\
y = ty_{V_{4}},\\
z = tz_{V_{4}},
\end{aligned}
\right.
\end{equation}
where $t$ is a positive real parameter. 
The intersection between the two spherical surfaces Eqs. \eqref{S1} and \eqref{S2} with the line Eq.
\eqref{L1} is the point ${\bf x}_0$ on the free surface with the following coordinates (where $z_{V_{4}}\neq 0$
by construction) 
\begin{equation}
\label{vault}
{\bf x}_0: \left \{ 
\begin{aligned}
x &= \frac{x_{V_{4}}}{z_{V_{4}}} z\\
y &= \frac{y_{V_{4}}}{z_{V_{4}}} z\\
z &= \frac{(-x_{C_{P}}^2+x_{C_{Q}}^2-y_{C_{P}}^2+y_{C_{Q}}^2-z_{C_{P}}^2
	+z_{C_{Q}}^2)z_{V_{4}}}{-2(x_{C_{P}}+x_{C_{Q}})x_{V_{4}}-2(y_{C_{P}}
	+y_{C_{Q}})y_{V_{4}}-2(z_{C_{P}}+z_{C_{Q}})z_{V_{4}}}.
\end{aligned}
\right.
\end{equation}
Hence, we define
\begin{equation}
h_2 = |{\bf x_0} - V_{4}|.
\end{equation}
This procedure can be repeated on all the lateral interfaces in order to calculate all the $h_i$ with $i=1,2,3,4$. 

The second and final step is to write down the volume of the solid (pyramidal frustum plus spherical vault) 
as a function of $h_P$, $h_T$ and $h_Q$ calculated as in Eq. \eqref{vault}.  
The volume of the pyramidal frustum is 
\begin{equation}
\label{eq:vpiramide}
\V_{\rm py}^{i} = \frac{\left(A_{\rm bottom}^{i}+A_{\rm top}^{i}+\sqrt{A_{\rm bottom}^{i}A_{\rm top}^{i}}\right)
	H_{\rm py}^{i}}{3} \qquad  i = P,Q,T,
\end{equation}
where $A_{\rm bottom}^{i}$ is the area of the $i$-th polygon on the dual tessellation, 
$A_{\rm top}^{i} = \frac{r+h}{r} A_{\rm bottom}^{i}$ is the area of the upper basis,  
and $H_{\rm py}^{i}$ is the height of the pyramidal frustum. \\
The volume of the spherical vault is nothing but the volume of the laterally cut spherical cap.
For the sake of simplicity, we consider here the spherical vault based on the triangle $N$. 
We use local coordinates with origin in $C_N$, the centroid of the triangle is in  
$ \tilde{N} = (0,0,h_T)$. \\
In terms of the local coordinates $(x',y',z')$, the volume of the spherical cap is 
$$\
\Omega' = \{ (x',y',z') \in \mathbb{R}^3 \, : (x')^2+(y')^2+(z')^2\leq R_T^2, \,  z'\geq h+h_T\}.
$$
and its measure is obtained by standard volume integration. 
We have to subtract to the measure of $\Omega'$ the volume of the slices obtained by prolongation
of the flat interface defined by the vertices $V_3V_7$ (and so on).
It is worth to remark  that the plane attached to $V_3V_7$ is not perpendicular to basis of the spherical cap.
So, first of all, the expression of the area of a circular segment of radius $\rho$
and chord $b$ is    
\begin{equation}
\mathcal{A}_{cir} = \rho^2 \left(\arcsin\left(\frac{b}{2\rho}\right)-\frac{b}{2\rho}\right).
\end{equation}
Here $\rho$ and $b$ are functions of the quote $z' \in [h+h_T, h_{\rm max}]$ and $h_{\rm max}$ 
has to be determined, see Fig. \ref{segmentocircolare}.
The value of $h_{\rm max}$ is fixed at the $z'$-level such that the homothetic projection of the inner triangle 
is circumscribed into the circumference. Namely, a plane at given $z'$ crosses the plane defined by 
$V_3V_7$ along a line, that we represent by its equation $ax'+by'+c=0$. To obtain the 
upper integration bounds in $z'$ we impose 
\begin{equation}
\label{condition}
\frac{|a\bar x'+b\bar y'+c|}{\sqrt{a^2+b^2}} = \sqrt{R_T^2-(z')^2},
\end{equation}
where $\bar x'=0$, $\bar y'=0$ and the left-hand-side of Eq. \eqref{condition} is exactly the radius $\rho$ 
of the circumference at fixed $z'$. \\
By solving \eqref{condition}, we get
$$
z'_{1,2} = \frac{-cd \pm \sqrt{c^2d^2-\left(a^2+b^2+c^2\right)(d^2-R_T^2a^2-R_T^2b^2)}}{c^2+a^2+b^2},
$$
where $a = 2.48-2.96h_{T}$, $b = 0$, $c = 1$ and $d = 0.42-h_{T}$ and $h_{\rm max}$ is the positive value, 
since it belongs to $\Omega'$.\\
\begin{figure}[h!]
	\centering
	\includegraphics[width=0.45\textwidth]{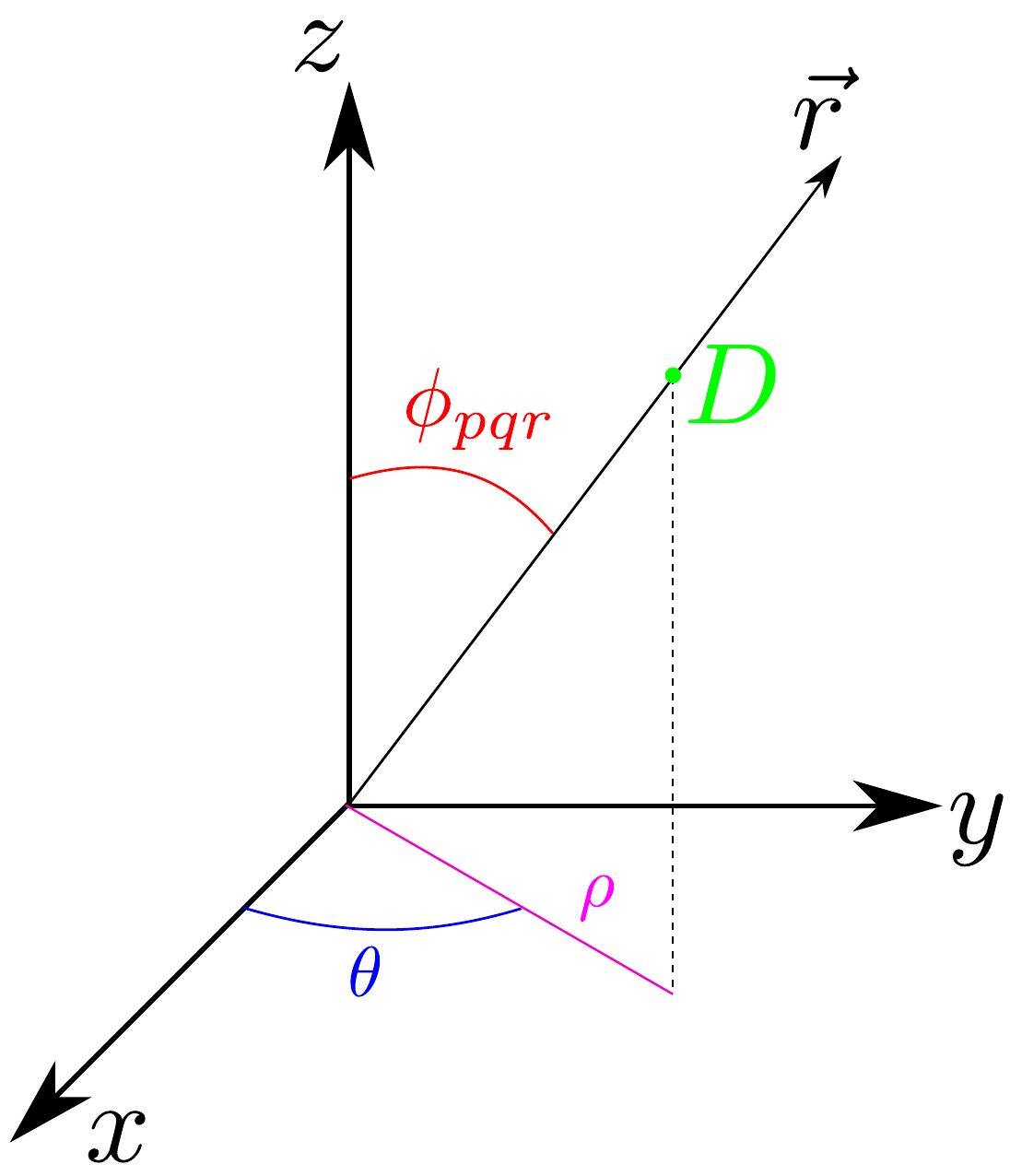}
	\caption{Geometrical sketch of the rotation around the axis $x$ by the angle $\phi_{pqr}$. 
		The coordinates $\rho$, $\theta$ and $\phi$ are the spherical ones set into 
		a Cartesian frame of reference $(x,y,z)$.}
	\label{fig:P1B}
\end{figure}
Finally, the volume of the bubble constructed on the triangle is given by
$$
\V_{T} = |\Omega'|- 3\int_{h+h_T}^{h_{\rm max}} \mathcal{A}_{cir} \, {\rm d} \alpha,
$$
where $|\cdot|$ denotes the measure of the volume of $\Omega'$. All the calculations are computed 
numerically by using the Newton's method with the software {\em Mathematica} 
11.3 (Wolfram Research,Champaign,IL, USA).
For the other polygons, the computation is similar up to a rotation which has to be done before 
computing the translation of the centre of the frame of reference, see Fig. \ref{fig:P1B}.

\section{Equations of balance of tensions}
\label{a2}
The detailed expressions of Eqs. \eqref{edge} and their geometrical representation are listed in Tables 
\ref{sotto}-\ref{sopra}. We introduce different superscripts to distinguish the different directions. 
The unit vectors denoting the direction of the force on the side edge is denoted by the symbol $\bm{t}$, 
$\bm{t}^{r}$ for the ones oriented in the radial direction, $\bm{t}^s$ on the free surface.

In the following tables, the unit vector normal to the free surface applied in the homothetical vertex of the dual Tammes' tessellation are not reported due to absence of space. We collect them below, {\em i.e.}                           
$$
\begin{aligned}
\bm{n}^s_{D} &= (-0.18+ 0.39 h_2 + 0.62 h_P, 0.53 + 
0.68 h_2, -0.4 - 0.64 h_2 - 0.13 h_P)\\
\bm{n}^s_{E} &= (0.55 + 0.39 h_2 - 0.3 h_P, 0.11 + 0.67 h_2 + 
0.58 h_P, -0.4 - 0.64 h_2 - 0.13 h_P)\\
\bm{n}^s_{A} &= (0.48 + 0.83 h_4 + 0.23 h_Q, -0.31 + 0.3 h_Q, 0.11 + 0.56 h_4 + 0.43 h_Q)\\
\bm{n}^s_{B} &=( 0.35 - 0.44 h_Q, 0, 0.46 + h_4 + 0.43 h_Q)\\
\bm{n}^s_{N} &=  (0.31+ 0.39 h_3, -0.53 - 0.67 h_3, -0.09 - 
0.64 h_3 - 0.53 h_T),
\end{aligned}
$$
where $h_P$, $h_Q$ and $h_T$ are the distance between the center of the sphere associated with the vault curvature and the centroid of the polygonal inner basis of the frustum, for more details see Appendix \ref{a3}.

\FloatBarrier
\begin{table}[h!]
	\centering
	\begin{tabular}{ |m{3.8cm}| m{4.6cm}| m{4.4cm} | m{2.5cm}|}
		\hline
		Adjacent polygons &Edges & Vectors & Equation  \\ \hline
		pentagon--pentagon
		\hspace{0.15cm}
		\centering
		\Includegraphics[width=0.85in]{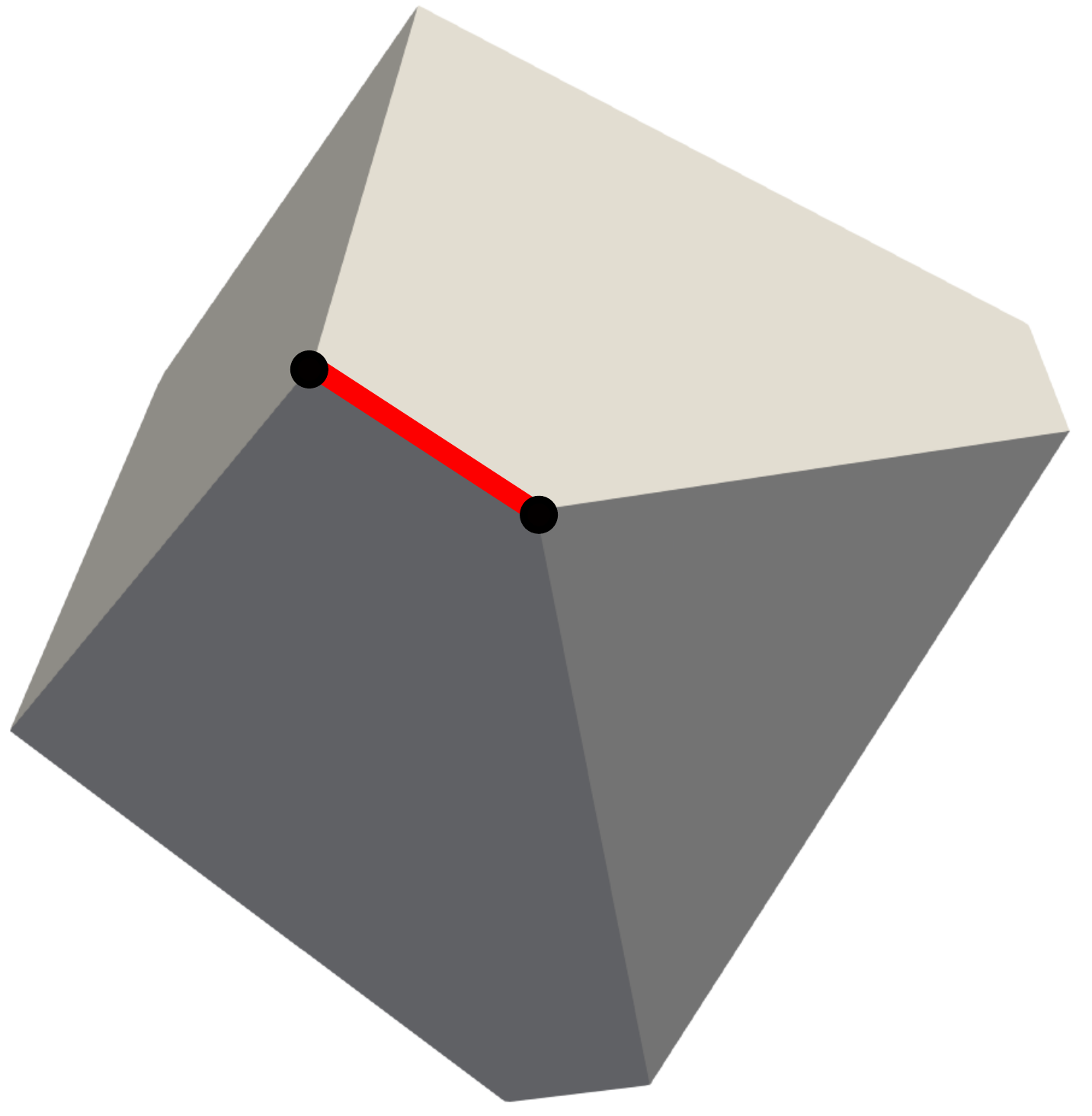}
		&
		\begin{minipage}{.3\textwidth}
			\Includegraphics[width=1.8in]{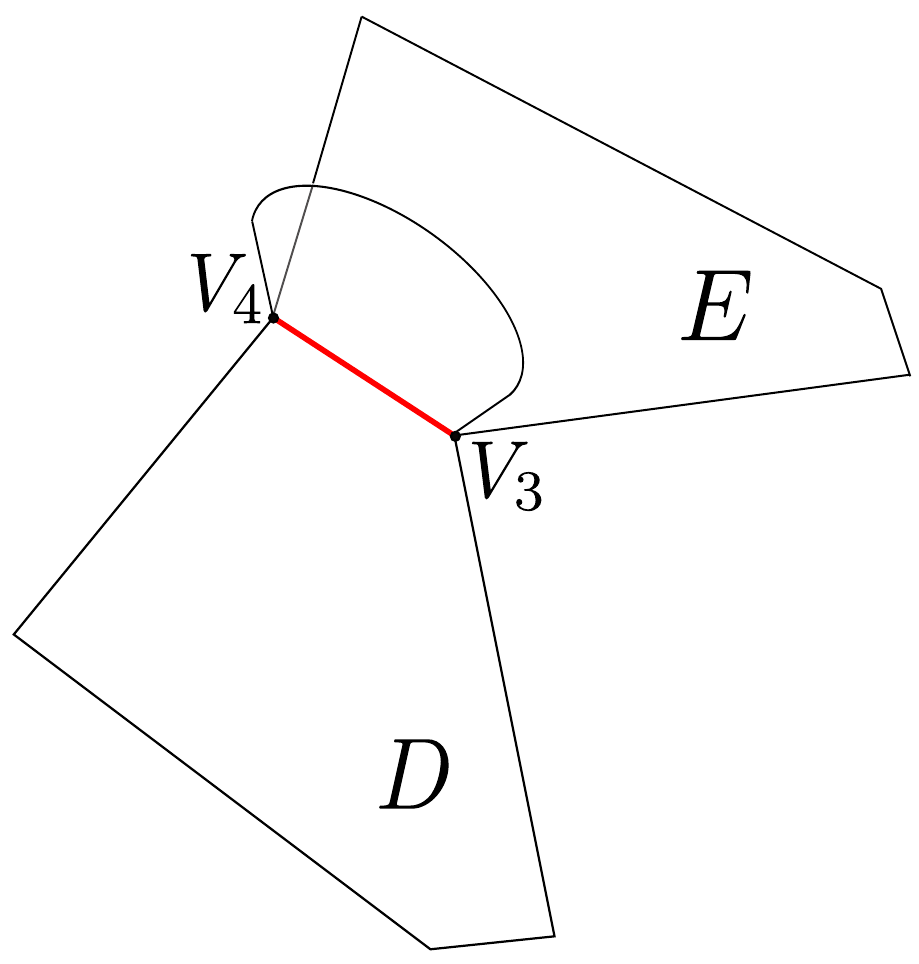} 
		\end{minipage}
		&
		$$
		\begin{footnotesize}
		\begin{aligned}
		{V}_3 &= (-0.3,-0.53,0.51)\\
		{V}_4 &= (-0.38,-0.66,0.2)\\
		\bm{n}_{D} &= (0.78,0,-0.16)\\
		\bm{n}_{E} &= (-0.38,0.67,-0.16)\\
		\bm{n}_{DE} &= (-0.3,0.13,0)\\
		\bm{c}_{3} & = (0.078,0.14,0.3)
		\end{aligned}
		\end{footnotesize}
		$$
		&
		$$
		\begin{aligned}
		&\tau_{P}\,\bm{t}_{\tiny{D}} \\
		+&\tau_{P}\,\bm{t}_{E}\\
		+&\tau_{PP}\,\bm{t}_{DE}\\
		&= 0
		\end{aligned}
		$$
		
		\\ \hline
		quadrilateral--pentagon
		\hspace{0.3cm}
		\centering
		\Includegraphics[width=0.85in]{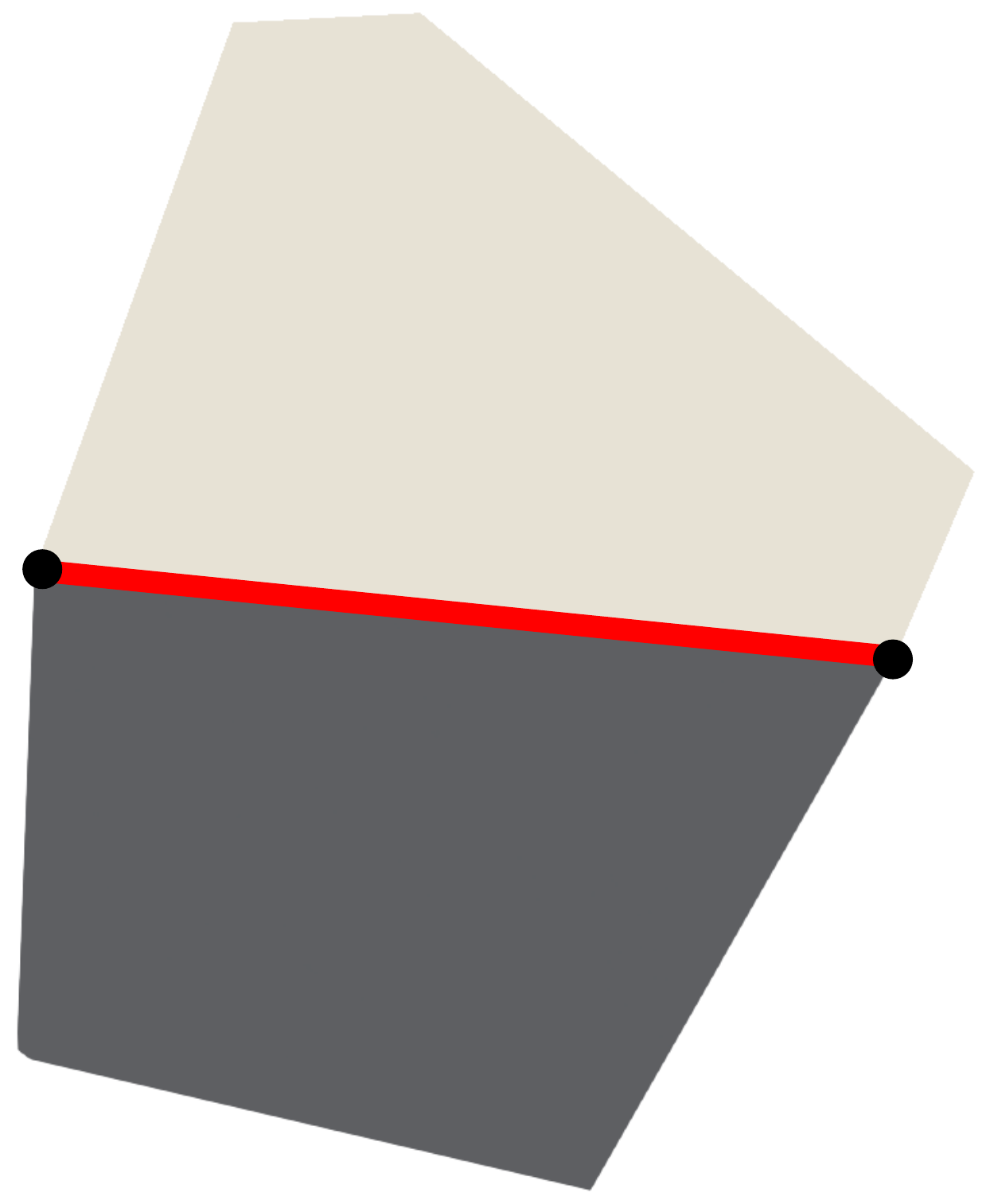}
		&
		\begin{minipage}{.3\textwidth}
			\Includegraphics[width=1.8in]{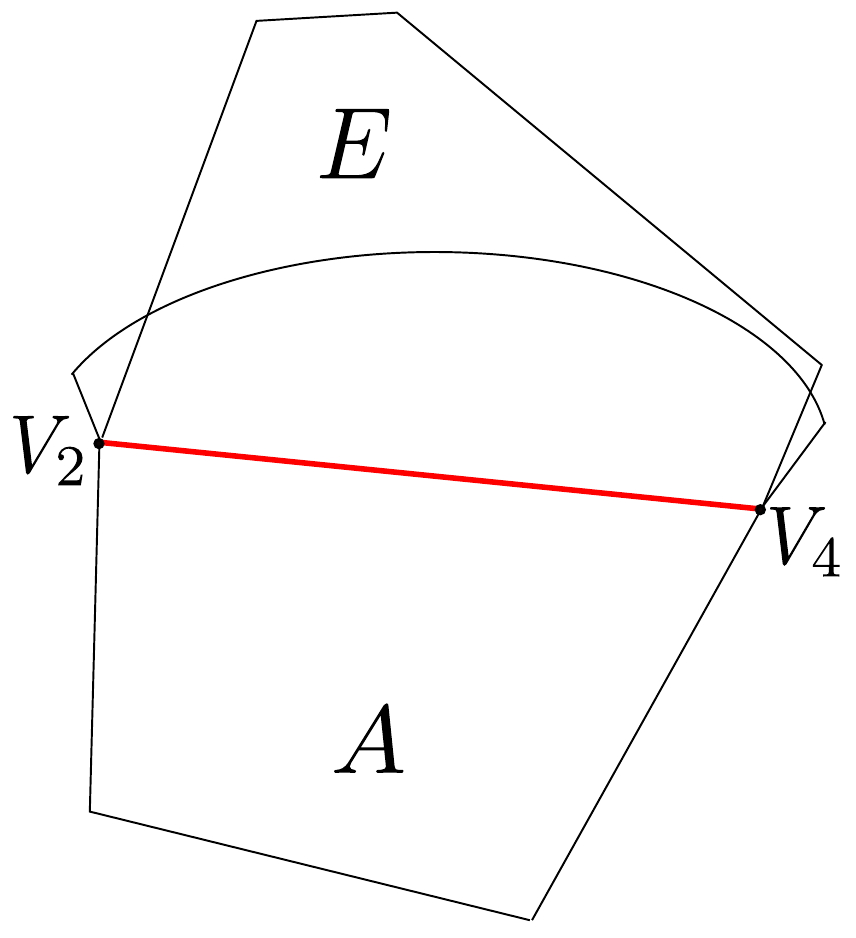}
		\end{minipage}
		&
		$$
		\begin{footnotesize}
		\begin{aligned}
		{V}_2 &= (-0.66,0,-0.45)\\
		{V}_4 &= (-0.38,-0.66,0.2)\\
		\bm{n}_{E} &= (0.78,0,-0.16)\\
		\bm{n}_{A} &= (0.28,0.5,0.55)\\
		\bm{n}_{AE} &= (-0.3,0.3,0.44)\\
		\bm{c}_{2} & = (-0.27,0.66,-0.65)
		\end{aligned}
		\end{footnotesize}
		$$
		&
		$$
		\begin{aligned}
		&\tau_{Q}\,\bm{t}_{A} \\
		+&\tau_{P}\,\bm{t}_{E}\\
		+&\tau_{PQ}\,\bm{t}_{AE}\\
		&= 0
		\end{aligned}
		$$
		
		\\ \hline
		quadrilateral--quadrilateral
		\centering
		\Includegraphics[width=0.85in]{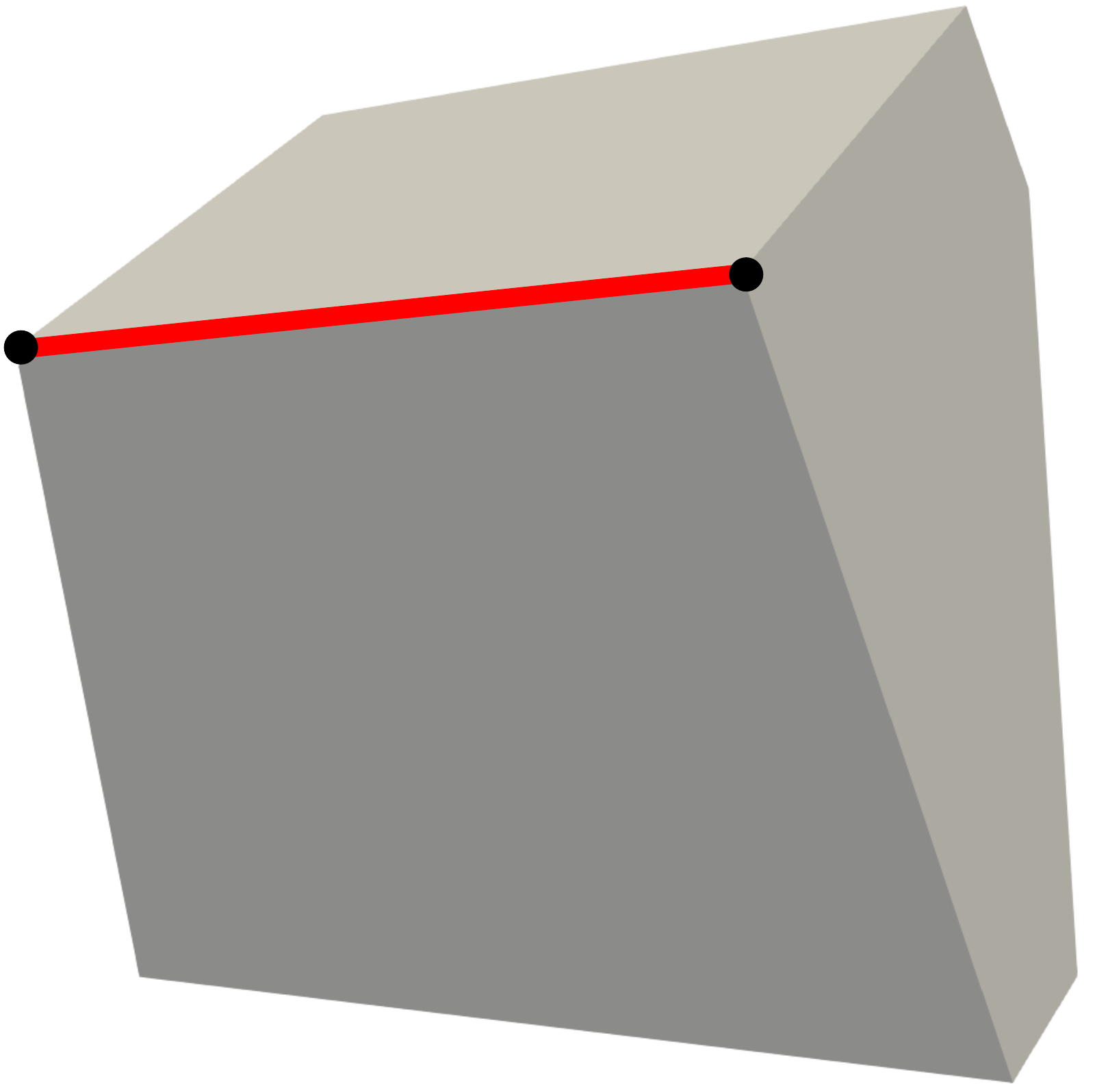}
		&
		\begin{minipage}{.3\textwidth}
			\Includegraphics[width=1.8in]{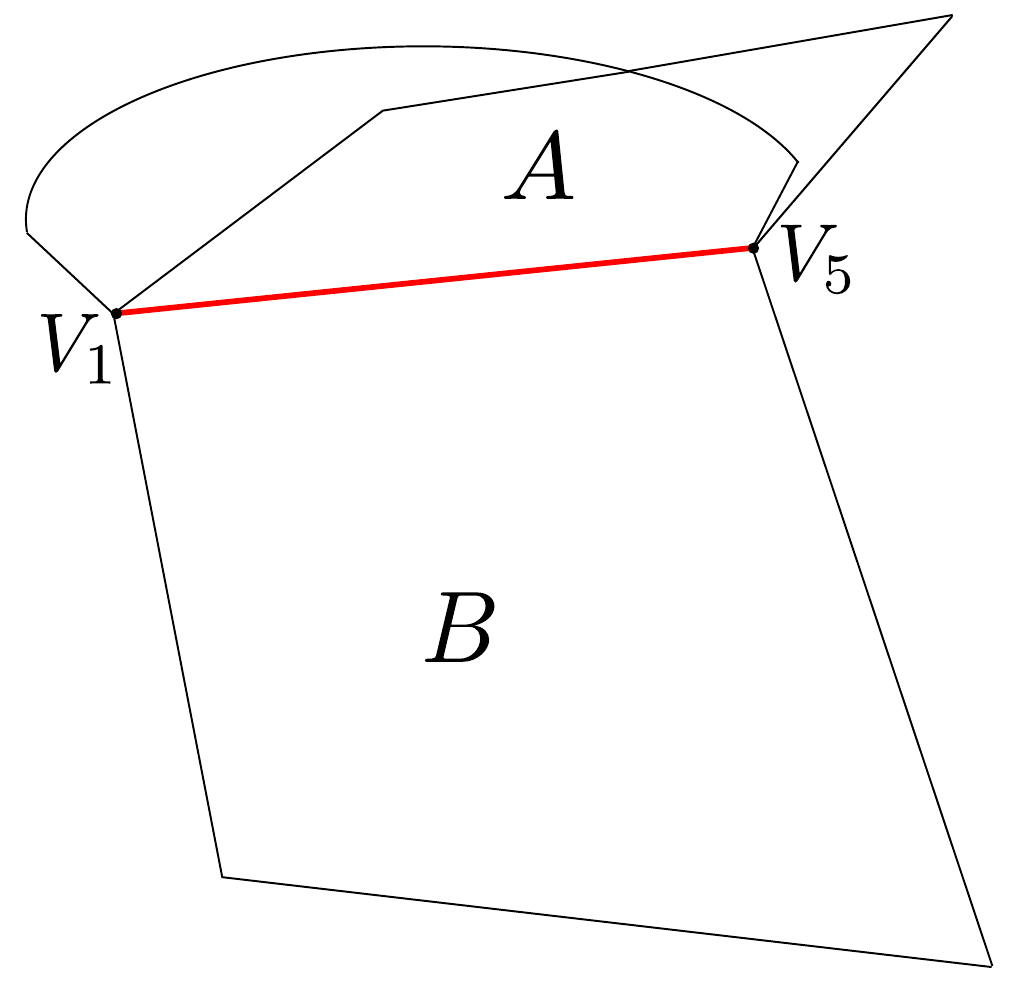} 
		\end{minipage}
		&
		$$
		\begin{footnotesize}
		\begin{aligned}
		{V}_1 &= (0,0,-0.79)\\
		{V}_5 &= (-0.65,0,-0.45)\\
		\bm{n}_{A} &= (0.28,0.5,0.55)\\
		\bm{n}_{B} &= (-0.57,0,0.55)\\
		\bm{n}_{AB} &= (0,0.52,0)\\
		\bm{c}_{1} & = (-0.66,0,0.35)
		\end{aligned}
		\end{footnotesize}
		$$
		&
		$$
		\begin{aligned}
		&\tau_{Q}\,\bm{t}_{A} \\
		+&\tau_{Q}\,\bm{t}_{B}\\
		+&\tau_{QQ}\,\bm{t}_{AB}\\
		&= 0
		\end{aligned}
		$$
		\\ \hline
		
		triangle--pentagon
		\centering
		\Includegraphics[width=0.85in]{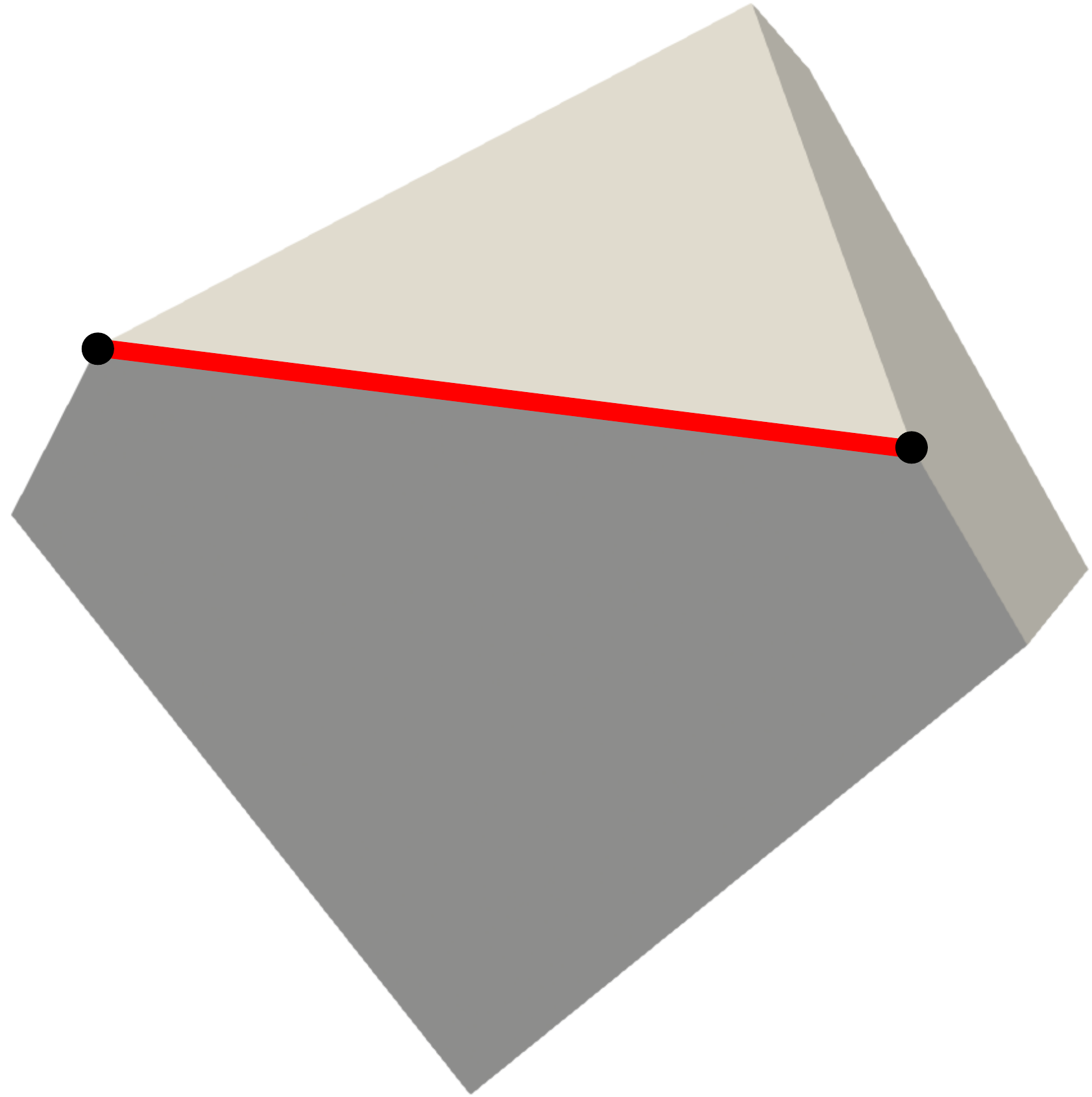}
		&
		\begin{minipage}{.3\textwidth}
			\Includegraphics[width=1.8in]{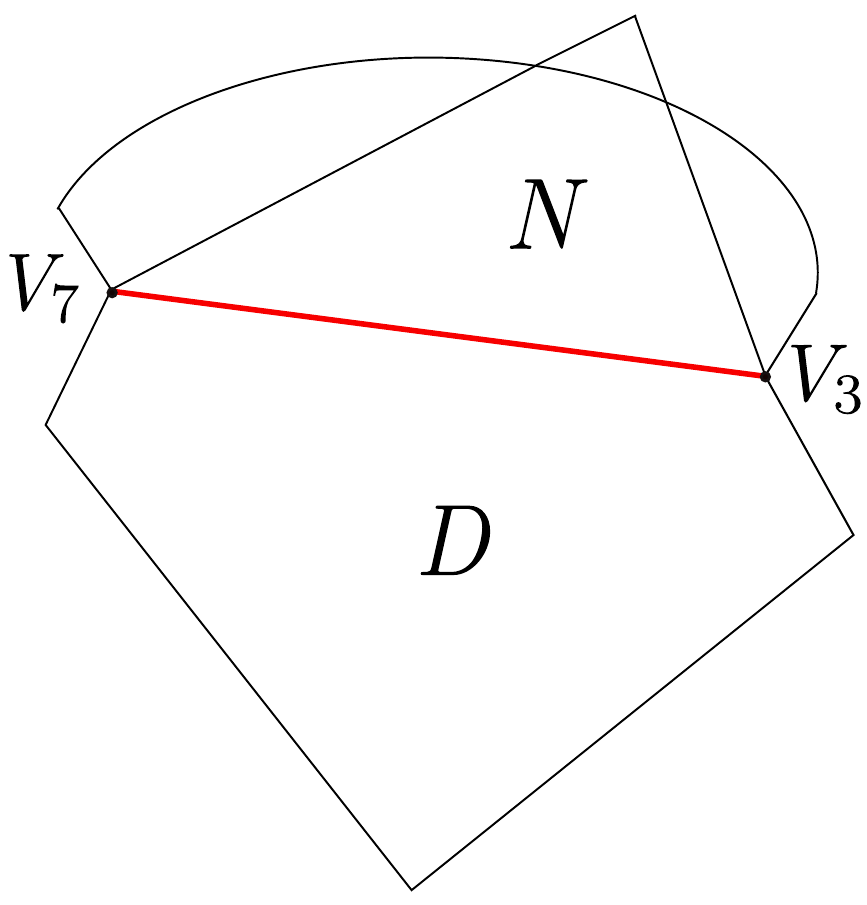}
		\end{minipage}
		&
		$$
		\begin{footnotesize}
		\begin{aligned}
		{V}_3 &= (-0.31,-0.53,0.51)\\
		{V}_7 &= (-0.31,0.53,0.51)\\
		\bm{n}_{D} &= (0.78,0,-0.16)\\
		\bm{n}_{N} &= (0,0,0.79)\\
		\bm{n}_{DN} &= (0.54,0,0.33)\\
		\bm{c}_{4} & = (0,1.06,0)
		\end{aligned}
		\end{footnotesize}
		$$
		&
		$$
		\begin{aligned}
		&\tau_{P}\,\bm{t}_{D} \\
		+&\tau_{T}\,\bm{t}_{N}\\
		+&\tau_{PT}\,\bm{t}_{DN}\\
		&= 0
		\end{aligned}
		$$
		\\ \hline
		
	\end{tabular}
	
	\caption{Edges on the tesselation}\label{sotto}
\end{table}
\FloatBarrier

\FloatBarrier
\begin{table}[h!]
	\centering  
	\begin{tabular}{ |m{2.8cm}| m{5cm} | m{5cm} | m{2.4cm}|}
		\hline
		Interaction& Edges & Angles & Equation  \\ \hline
		quadrilateral--quadrilateral--quadrilateral
		\centering
		\Includegraphics[width=0.85in]{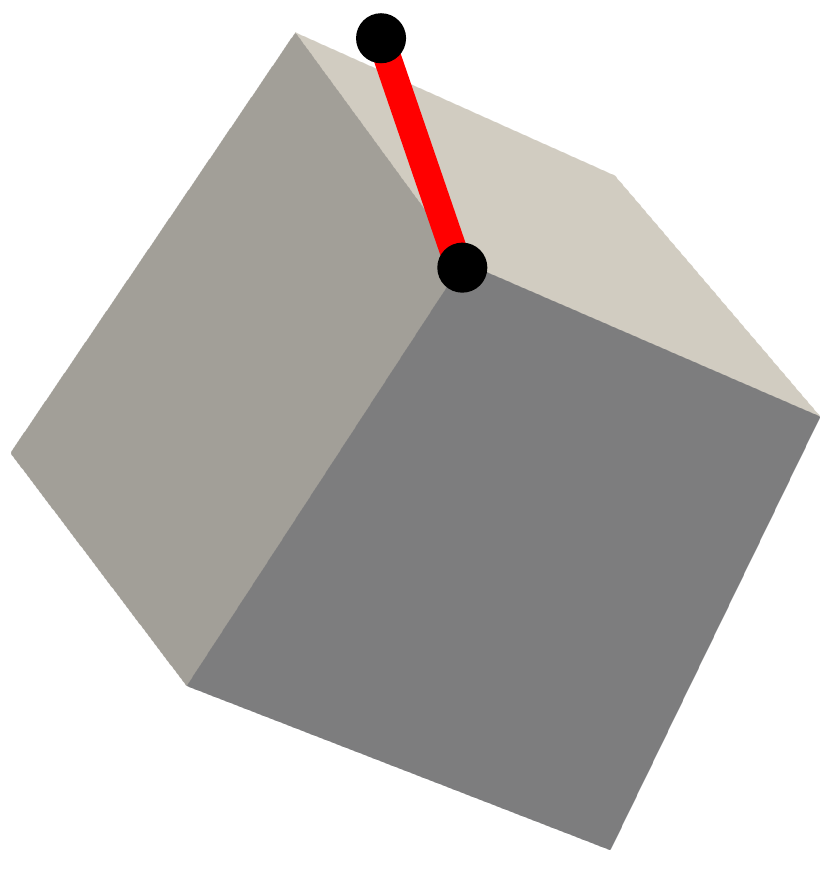}&
		\begin{minipage}{.3\textwidth}
			\Includegraphics[width=2in]{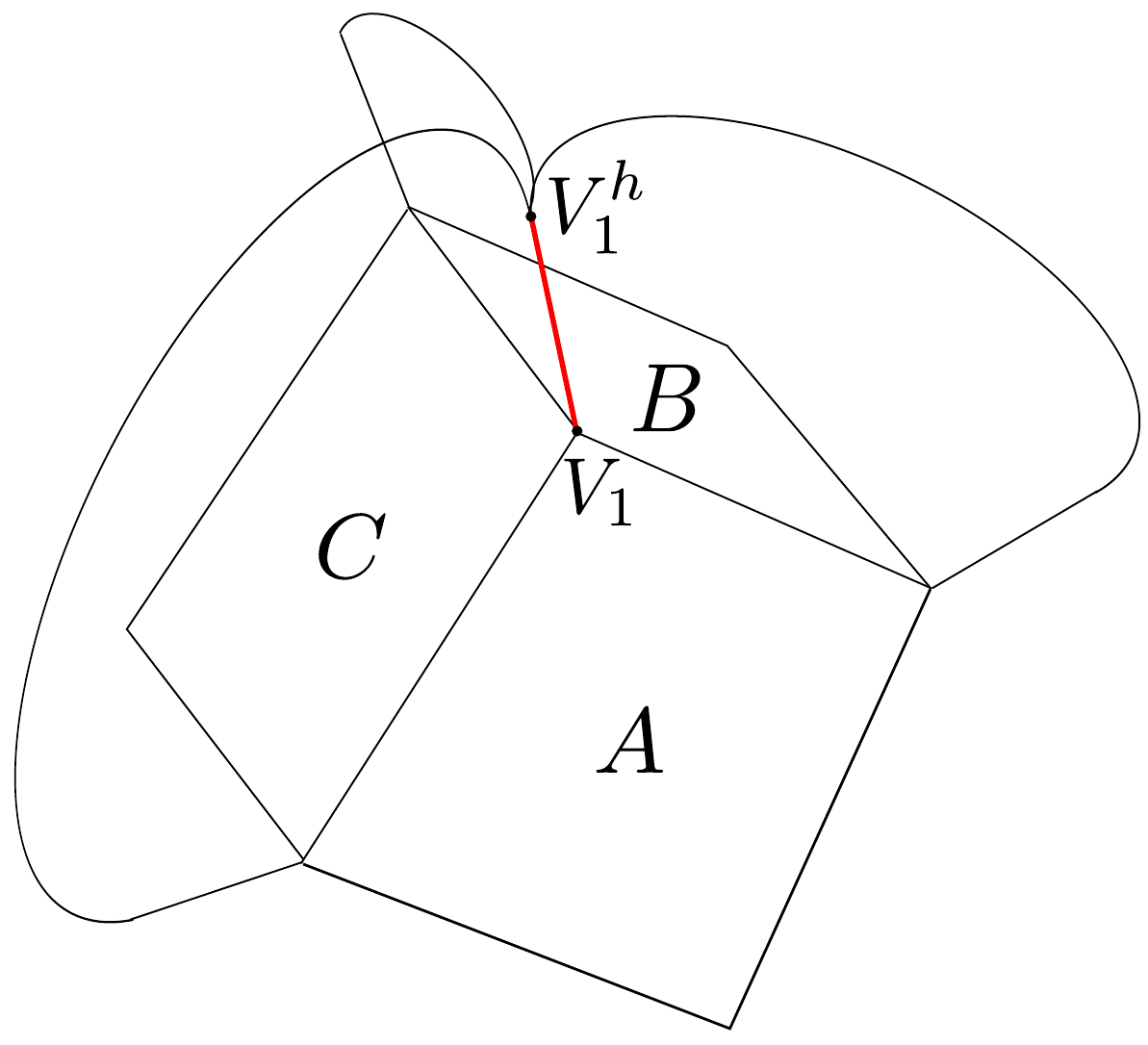}
		\end{minipage}
		&
		$$
		\begin{footnotesize}
		\begin{aligned}
		V_{1}^h &= (0,0,-0.79-h_4)\\
		\bm{n}_{AB} &=  (0, 0.52, 0)\\
		\bm{n}_{BC} &= (0.45, -0.26, 0)\\
		\bm{n}_{CA} &= (-0.45, -0.26, 0)\\
		\bm{c}_{5} & = (0,0,h_4)
		\end{aligned}
		\end{footnotesize}
		$$
		&
		$$
		\begin{aligned}
		&\tau_{QQ}\,\bm{t}^{r}_{AB} \\
		+&\tau_{QQ}\,\bm{t}^{r}_{BC}\\
		+&\tau_{QQ}\,\bm{t}^{r}_{CA}\\
		&= 0
		\end{aligned}
		$$
		
		
		\\ \hline
		pentagon--pentagon--triangle
		\centering
		\Includegraphics[width=0.85in]{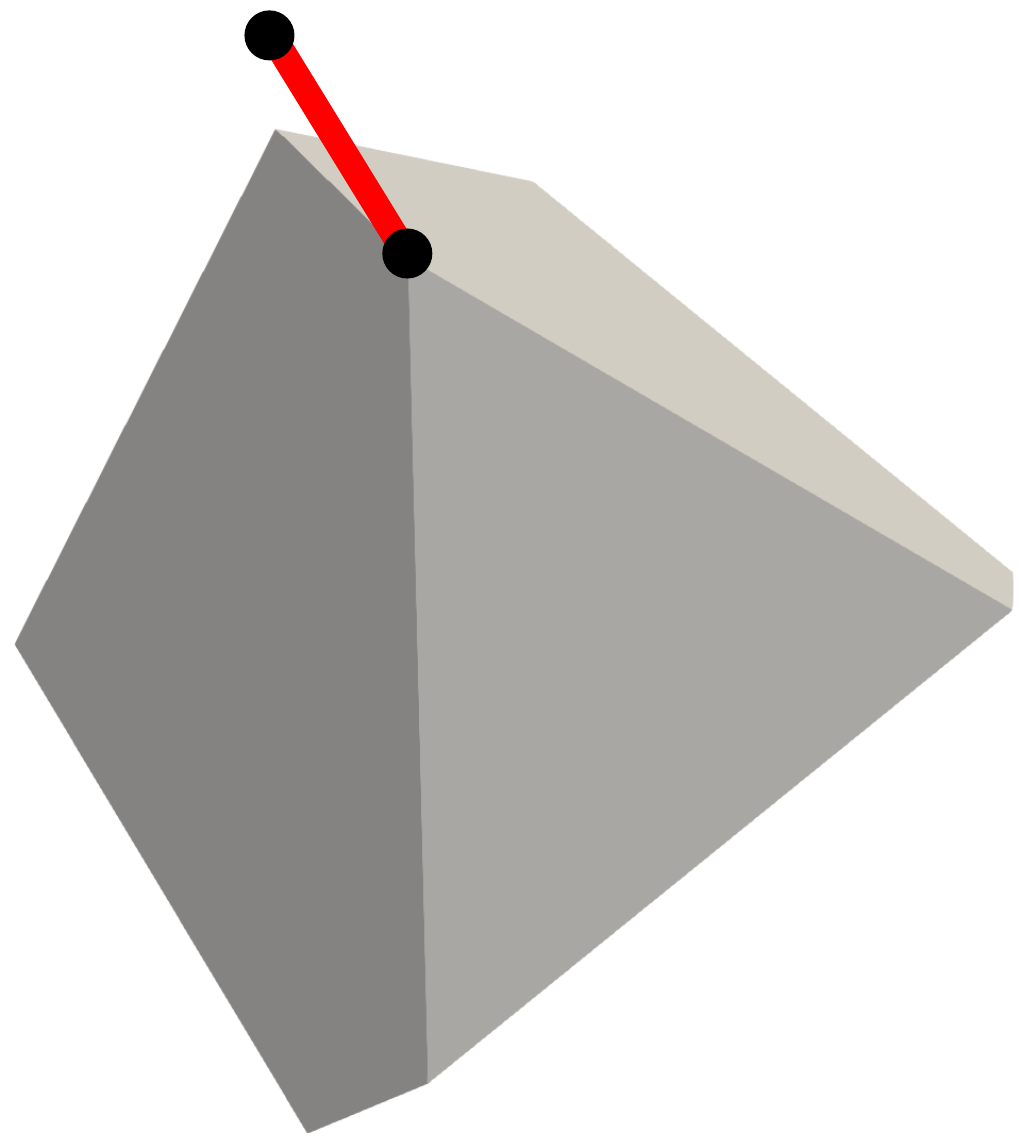}&
		\begin{minipage}{.3\textwidth}
			\Includegraphics[width=2in]{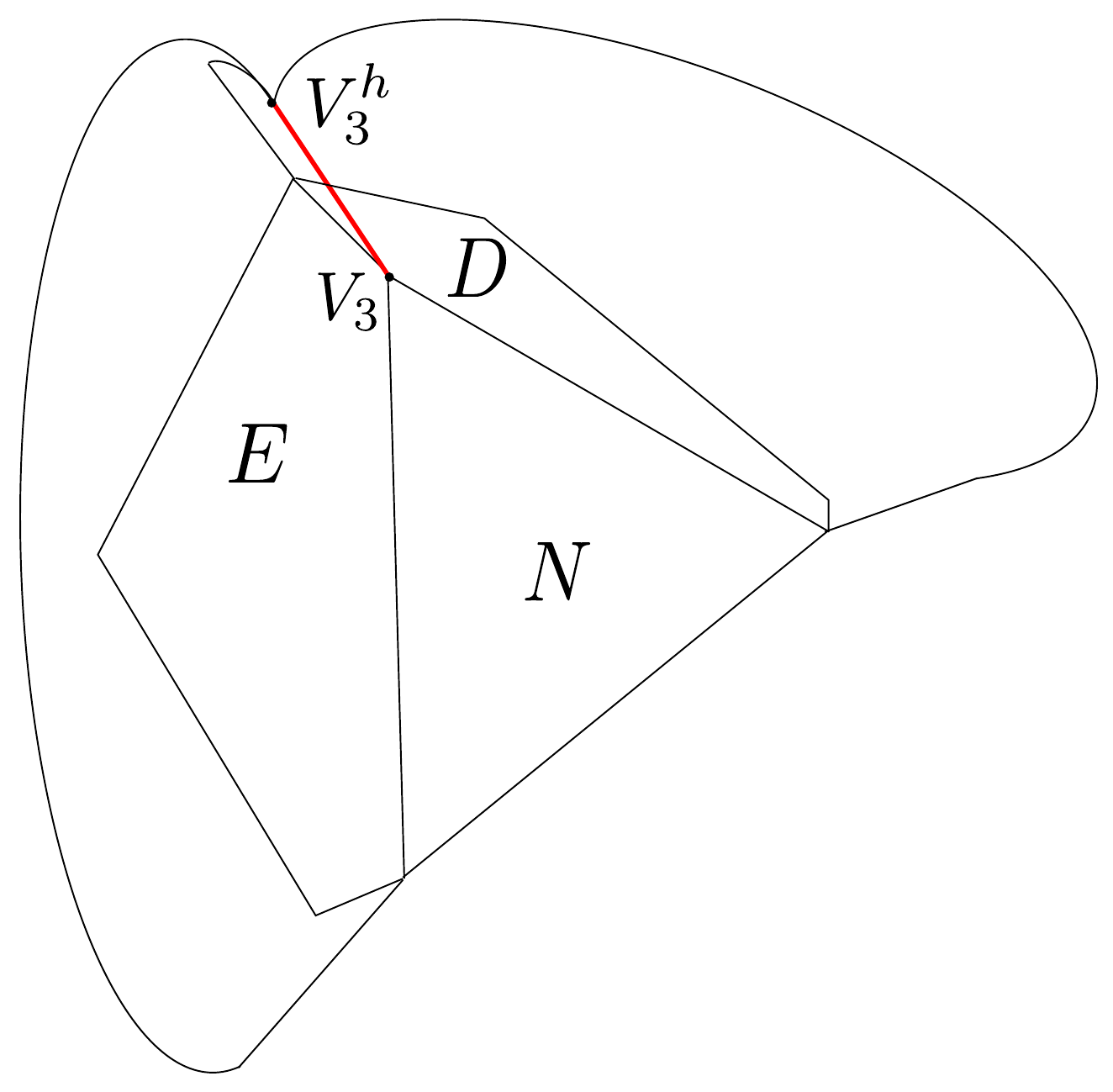}
		\end{minipage}
		&
		$$
		\begin{footnotesize}
		\begin{aligned}
		V_{3}^h &= (-\hspace{0.08cm}0.31 - 0.39 h_3,\\
		&\hspace{0.55cm}-0.53 - 0.67 h_3,\\
		&\hspace{1cm} 0.51 + 0.64 h_3)\\
		\bm{n}_{DE} &= (-0.3, 0.13, 0) \\
		\bm{n}_{DN} &= (-0.27, 0.47, 0.33)\\
		\bm{n}_{EN} &= (-0.54, 0, 0.33)\\
		\bm{c}_{7} & = ( 0.39h_3, 0.67 h_3, - 0.64 h_3)
		\end{aligned}
		\end{footnotesize}
		$$
		&
		$$
		\begin{aligned}
		&\tau_{PT}\,\bm{t}^{r}_{DN} \\
		+&\tau_{PT}\,\bm{t}^{r}_{EN}\\
		+&\tau_{PP}\,\bm{t}^{r}_{DE}\\
		&= 0
		\end{aligned}
		$$
		
		\\ \hline
		quadrilateral--quadrilateral--pentagon
		\centering
		\Includegraphics[width=0.85in]{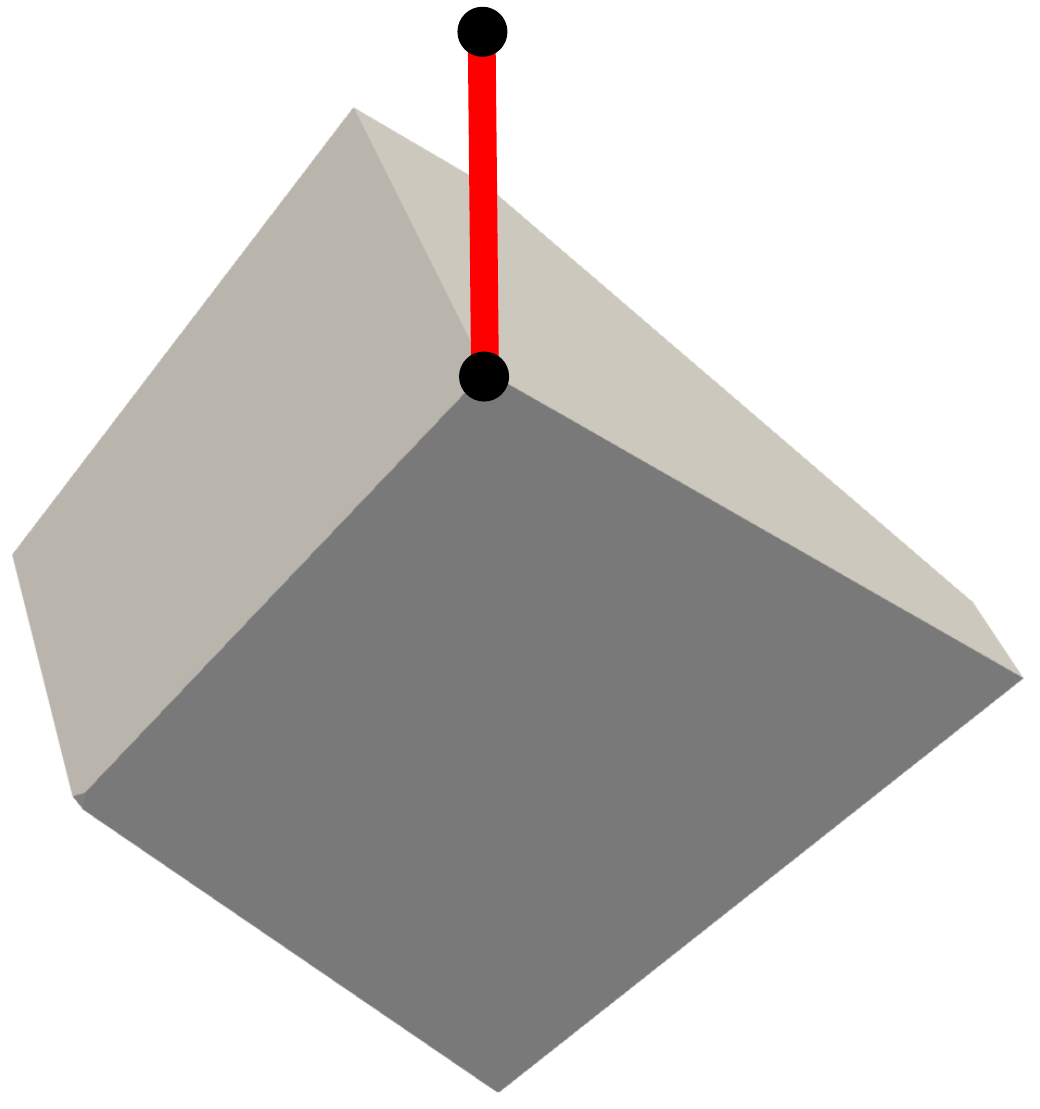}&
		\begin{minipage}{.3\textwidth}
			\Includegraphics[width=1.8in]{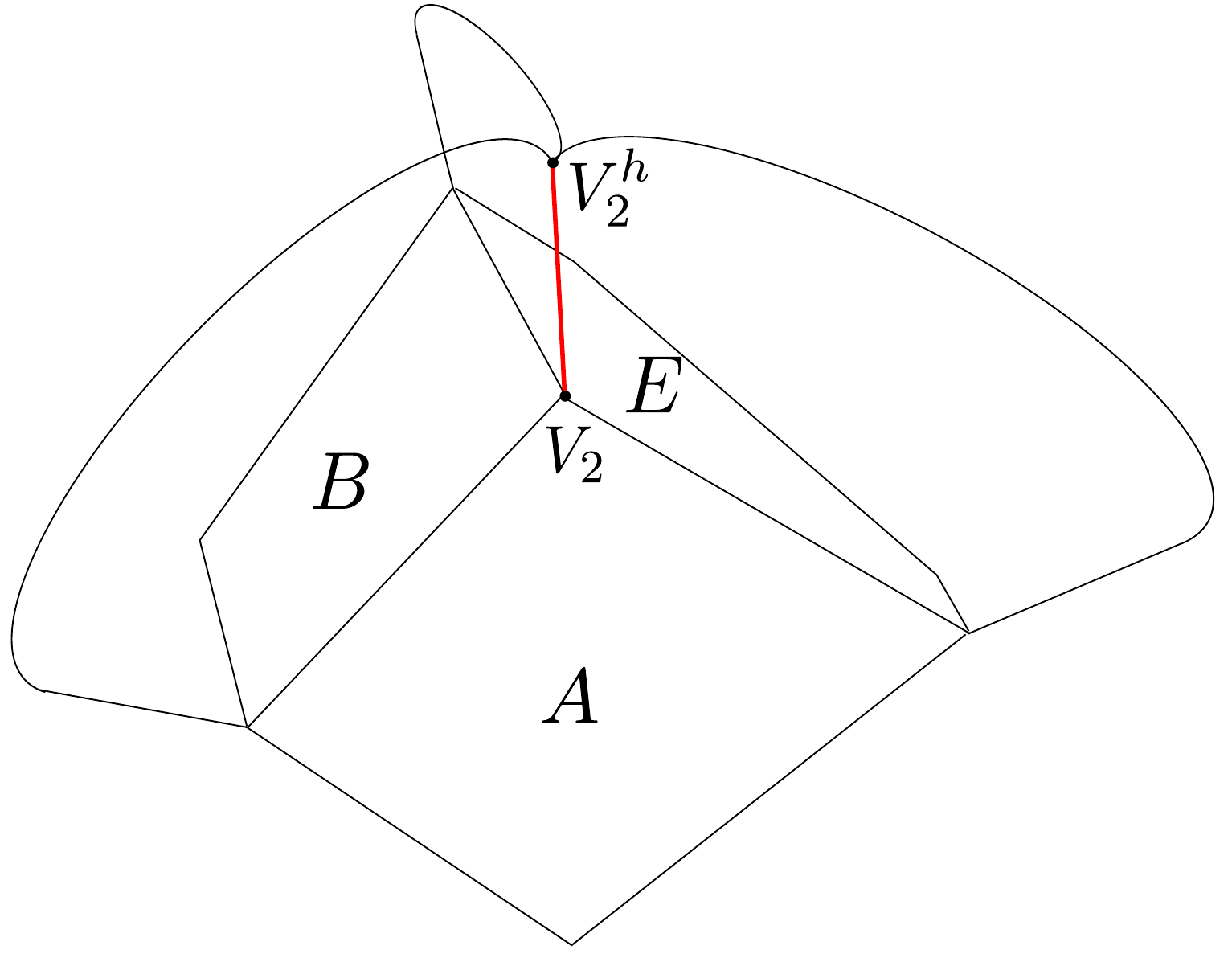}
		\end{minipage}
		&
		$$
		\begin{footnotesize}
		\begin{aligned}
		{V}_2^h &= (-\hspace{0.07cm}0.66 - 0.83 h_1, \\
		&\hspace{0.95cm}0,\\
		&\hspace{0.53cm}-0.45 - 0.56 h_1)\\
		\bm{n}_{AB} &= (0, 0.52, 0)\\
		\bm{n}_{AE} &= (-0.3, 0.31, 0.44)\\
		\bm{n}_{BE} &= (0.3, 0.31, -0.44)\\
		\bm{c}_{6} & = ( 0.83 h_1, 0, 0.56 h_1) 
		\end{aligned}
		\end{footnotesize}
		$$
		&
		$$
		\begin{aligned}
		&\tau_{QQ}\,\bm{t}^{r}_{AB} \\
		+&\tau_{PQ}\,\bm{t}^{r}_{AE}\\
		+&\tau_{PQ}\,\bm{t}^{r}_{BE}\\
		&= 0
		\end{aligned}
		$$
		\\ \hline
		
		pentagon--pentagon--quadrilateral
		\centering
		\Includegraphics[width=0.85in]{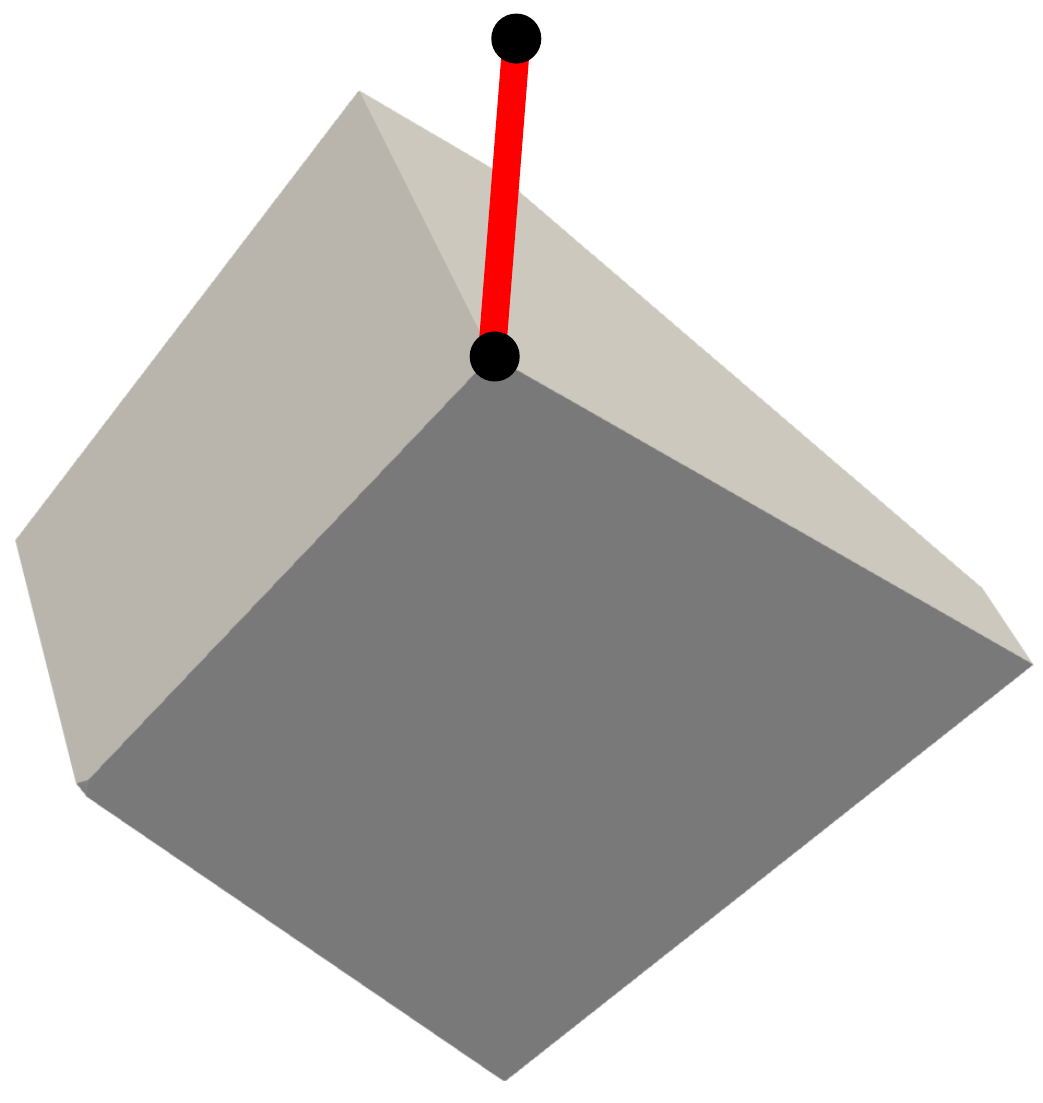}&
		\begin{minipage}{.3\textwidth}
			\Includegraphics[width=1.8in]{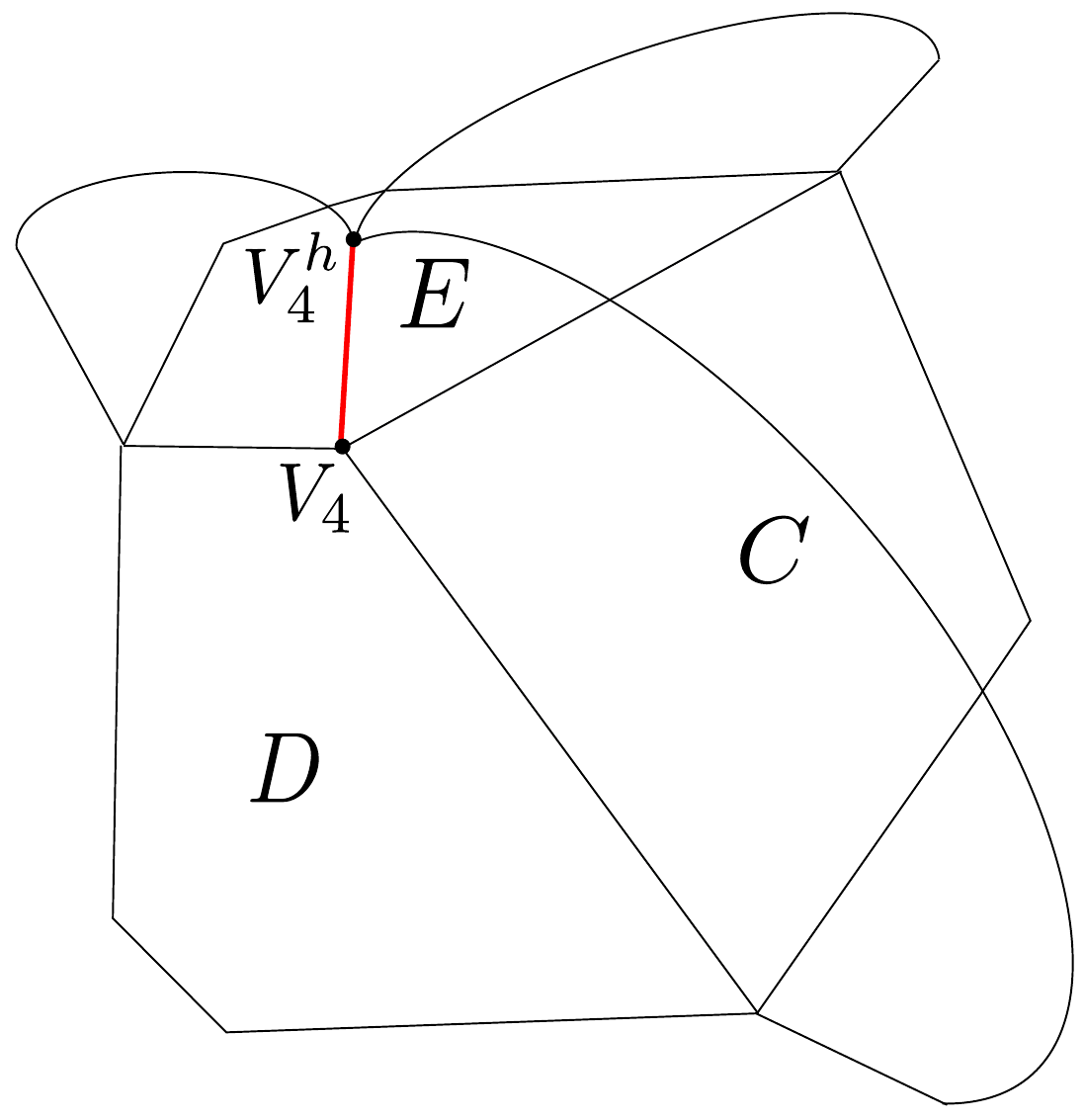} 
		\end{minipage}
		&
		$$
		\begin{footnotesize}
		\begin{aligned}
		{V}_{4}^h &= (-\hspace{0.07cm}0.38 - 0.48 h_1,\\
		&\hspace{0.55cm}-0.67 - 0.84 h_1,\\
		&\hspace{1cm}0.20 + 0.25 h_1)\\
		\bm{n}_{DE} &=  (-0.23, 0.13, 0)\\
		\bm{n}_{CD} &= (-0.3, 0.31, 0.44)\\
		\bm{n}_{CE} &=(0.41, -0.17, 0.44)\\
		\bm{c}_{8} & = (0.48 h_1, 0.84 h_1, - 0.25 h_1)
		\end{aligned}
		\end{footnotesize}
		$$
		&
		$$
		\begin{aligned}
		&\tau_{PP}\,\bm{t}^{r}_{DE} \\
		+&\tau_{PQ}\,\bm{t}^{r}_{CD}\\
		+&\tau_{PQ}\,\bm{t}^{r}_{CE}\\
		&= 0
		\end{aligned}
		$$
		
		\\ \hline
		
	\end{tabular}
	
	\caption{Radial edges}\label{radiale}
\end{table}
\FloatBarrier

\FloatBarrier
\begin{table}[h!]
	\centering 
	\begin{tabular}{|m{2.5cm}| m{4.5cm} | m{5cm} | m{2.5cm}|}
		\hline
		Interaction & Edges & Angles & Equation  \\ \hline
		pentagon--pentagon&
		\begin{minipage}{.3\textwidth}
			\Includegraphics[width=1.8in]{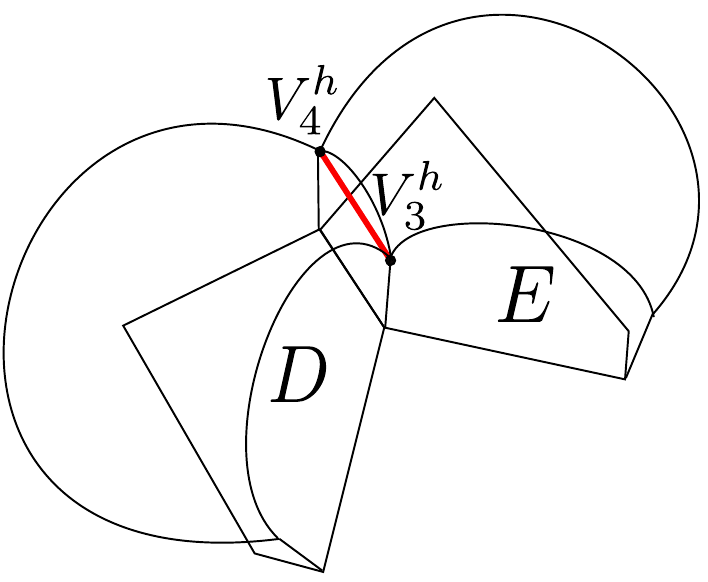}
		\end{minipage}
		&
		$$
		\begin{footnotesize}
		\begin{aligned}
		\bm{n}_{DE} &= (-0.23, 0.13, 0)\\
		\bm{c}_{11} & = (\hspace{0.07cm}0.078 + 0.1 h_3,\\
		&\hspace{0.7cm}0.14 + 0.17 h_3, \\
		&\hspace{0.7cm}0.3 + 0.38 h_3)
		\end{aligned}
		\end{footnotesize}
		$$
		&
		$$
		\begin{aligned}
		&\tau_{P}^s\,\bm{t}_{D}^s\\
		+&\tau_{P}^s\,\bm{t}_{E}^s\\
		+&\tau_{PP}\,\bm{t}^{s}_{DE}\\
		&= 0
		\end{aligned}
		$$
		
		\\ \hline
		quadrilateral--pentagon&
		\begin{minipage}{.3\textwidth}
			\Includegraphics[width=1.8in]{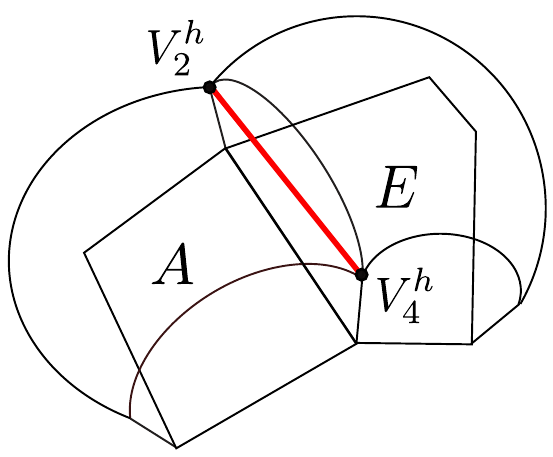}
		\end{minipage}
		&
		$$ 
		\begin{footnotesize}
		\begin{aligned}
		\bm{n}_{AE} &= (0.41, -0.17, 0.44)\\
		\bm{c}_{10} & = (\hspace{0.07cm}-0.27 - 0.34h_1,\\
		&\hspace{0.95cm}0.67 + 0.84 h_1, \\
		&\hspace{0.53cm}-0.65- 0.82 h_1)
		\end{aligned}
		\end{footnotesize}
		$$
		&
		$$
		\begin{aligned}
		&\tau_{P}^s\,\bm{t}_{E}^s\\
		+&\tau_{Q}^s\,\bm{t}_{A}^s\\
		+&\tau_{PQ}\,\bm{t}^{s}_{AE}\\
		&= 0
		\end{aligned}
		$$
		
		\\ \hline
		quadrilateral--quadrilateral&
		\begin{minipage}{.3\textwidth}
			\Includegraphics[width=1.8in]{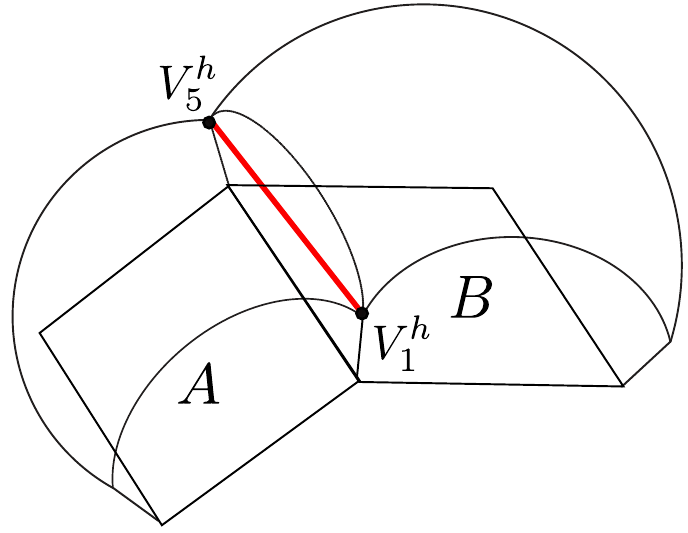} 
		\end{minipage}
		&
		$$
		\begin{footnotesize}
		\begin{aligned}
		\bm{n}_{AB} &= (0,0.52,0)\\
		\bm{c}_{9} & = (\hspace{0.07cm} -0.66 - 0.82 h_4,\\
		&\hspace{0.96cm}0, \\
		&\hspace{0.97cm} 0.35 + 0.44 h_4)
		\end{aligned}
		\end{footnotesize}
		$$
		&
		$$
		\begin{aligned}
		&\tau_{Q}^s\,\bm{t}_{A}^s\\
		+&\tau_{Q}^s\,\bm{t}_{B}^s\\
		+&\tau_{QQ}\,\bm{t}^{s}_{AB}\\
		&= 0
		\end{aligned}
		$$
		\\ \hline
		
		pentagon--triangle&
		\begin{minipage}{.3\textwidth}
			\Includegraphics[width=1.8in]{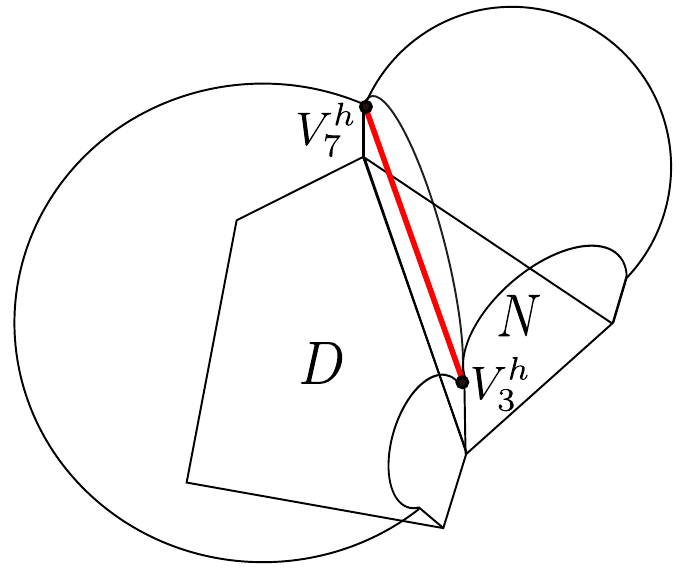}
		\end{minipage}
		&
		$$
		\begin{footnotesize}
		\begin{aligned}
		\bm{n}_{DN} &= (-0.54,0,0.33)\\
		\bm{c}_{12} & = (0, -1.06 - 1.34 h_3, 0)
		\end{aligned}
		\end{footnotesize}
		$$
		&
		$$
		\begin{aligned}
		&\tau_{P}^s\,\bm{t}_{D}^s\\
		+&\tau_{T}^s\,\bm{t}_{N}^s\\
		+&\tau_{PT}\,\bm{t}^{s}_{DN}\\
		&= 0
		\end{aligned}
		$$
		
		\\ \hline
		
	\end{tabular}
	
	\caption{Edges on the free surface. The expressions of $\bm{n}_i^s$ are reported at the beginning of this Appendix.
	}\label{sopra}
\end{table}
\FloatBarrier

\end{appendices}

\end{document}